\documentclass[reprint,aps,prd,superscriptaddress,nofootinbib,
               longbibliography]{revtex4-2}

\usepackage{graphicx}
\usepackage{amsmath,amssymb,bm}
\usepackage{xcolor}
\usepackage[colorlinks=true,linkcolor=blue,citecolor=blue,
            urlcolor=blue]{hyperref}

\colorlet{BLUE}{blue}
\newcommand{\rev}[1]{#1}
\definecolor{chgcol}{rgb}{0.55,0.00,0.55}
\newcommand{\chg}[1]{#1}
\graphicspath{{figs/}}

\newcommand{\Msun}{M_\odot}
\newcommand{\hc}{h_c}
\newcommand{\Ayr}{A_{\rm yr}}
\newcommand{\Mbh}{M_{\rm BH}}
\newcommand{\dkh}{\Delta_{\rm KH}}
\newcommand{\sdex}{\sigma_{\rm dex}}
\newcommand{\thard}{\tau_{\rm hard}}
\newcommand{\rhob}{\rho_\times}
\newcommand{\et}{e_t}
\newcommand{\sashimi}{{\sc sashimi}}
\newcommand{\holodeck}{{\sc holodeck}}

\newcommand{\AyrDMX}{2.4^{+0.7}_{-0.6}\times10^{-15}}
\newcommand{\AyrCNM}{2.1^{+0.6}_{-0.5}\times10^{-15}}
\newcommand{\gammaDMX}{3.2 \pm 0.6}
\newcommand{\gammaCNM}{3.5^{+0.7}_{-0.6}}

\newcommand{\AyrPLaspub}{2.51^{+0.59}_{-0.61}\times10^{-15}}
\newcommand{\gammaPLaspub}{3.07^{+0.63}_{-0.56}}
\newcommand{\AyrPLfid}{2.29^{+0.55}_{-0.52}\times10^{-15}}
\newcommand{\gammaPLfid}{3.31^{+0.73}_{-0.64}}
\newcommand{\capturedamp}{72\%}
\newcommand{\capturedgam}{81\%}
\newcommand{\dkhT}{+0.31^{+0.20}_{-0.23}}
\newcommand{\sdexT}{0.48^{+0.23}_{-0.30}}
\newcommand{\thardT}{3.1 \pm 2.0}
\newcommand{\etT}{0.67^{+0.16}_{-0.35}}
\newcommand{\lrhobT}{+0.52^{+0.71}_{-1.14}}
\newcommand{\dkhval}{+0.31^{+0.20}_{-0.23}}
\newcommand{\dkhsigT}{1.3$\sigma$}
\newcommand{\pdkhT}{91\%}
\newcommand{\rdkhtau}{+0.26}
\newcommand{\rdkhsig}{-0.72}
\newcommand{\rdkhet}{+0.41}
\newcommand{\fstallpure}{38\%}
\newcommand{\fnotcoalT}{$46 \pm 6\%$}
\newcommand{\dkhB}{+0.25^{+0.20}_{-0.24}}
\newcommand{\etB}{0.64^{+0.18}_{-0.36}}
\newcommand{\dkhdiffdel}{0.06}

\newcommand{\dkhshiftLC}{-0.07}
\newcommand{\dkhaspub}{+0.42^{+0.23}_{-0.24}}
\newcommand{\etaspub}{0.79^{+0.08}_{-0.21}}

\newcommand{\dkhshiftaspub}{+0.11}
\newcommand{\pdkhaspub}{97\%}

\newcommand{\dkhzOne}{-0.00^{+0.96}_{-0.94}}

\newcommand{\dkhMM}{+0.51^{+0.19}_{-0.24}}

\newcommand{\dkhMH}{+0.59^{+0.19}_{-0.24}}

\newcommand{\fwandgal}{10\%}
\newcommand{\fwandcos}{21\%}
\newcommand{\fwandclus}{62\%}
\newcommand{\fwandcross}{2\times10^{12}}
\newcommand{\fwandslope}{1.2}
\newcommand{\Nwandmw}{0.5}
\newcommand{\fsatbk}{26\%}
\newcommand{\fwandgalbk}{5.6\%}
\newcommand{\fwandcosbk}{15\%}
\newcommand{\fwandslopeN}{1.0}
\newcommand{\fwandthirteen}{29\%}
\newcommand{\fwandthirteenbk}{20\%}
\newcommand{\fstalltransbk}{13\%}
\newcommand{\fstallpureth}{64\%}
\newcommand{\fstallpermth}{49\%}
\newcommand{\fstalltransth}{15\%}
\newcommand{\fsatbkth}{46\%}
\newcommand{\fstallbkth}{27\%}
\newcommand{\fstalltransbkth}{19\%}
\newcommand{\fstallperm}{21\%}
\newcommand{\fstalltrans}{17\%}
\newcommand{\xiperm}{0.04}
\newcommand{\fstallbk}{13\%}

\newcommand{\tiltwall}{-0.20}
\newcommand{\tiltbk}{-1.44}
\newcommand{\Ndisc}{140}
\newcommand{\Ndischi}{160}
\newcommand{\Ndiscvol}{290}
\newcommand{\cvscatlow}{0.03}
\newcommand{\cvscathigh}{0.18}
\newcommand{\cvratiomax}{0.50}
\newcommand{\cvbinmax}{3}
\newcommand{\nbinfid}{13}
\newcommand{\cvinflmax}{12\%}

\newcommand{\dAtinkfit}{-2.7\%}
\newcommand{\dAsiafit}{-6.5\%}
\newcommand{\dAstbr}{-1.9\%}

\newcommand{\dAdelay}{-11.2\%}
\newcommand{\dAdelayabs}{11.2\%}
\newcommand{\dAfrozen}{-9.5\%}
\newcommand{\dAfrozenabs}{9.5\%}

\newcommand{\dAorbit}{1.4\%}
\newcommand{\dAkernel}{-3.4\%}
\newcommand{\dAkernelabs}{3.4\%}
\newcommand{\dAlc}{+14\%}
\newcommand{\dAlcabs}{14\%}
\newcommand{\dAkernelgain}{4}
\newcommand{\Ayrtidal}{1.45\times 10^{-15}}
\newcommand{\Ayrbk}{1.64\times 10^{-15}}
\newcommand{\fnotfidT}{39\%}
\newcommand{\fnotfidB}{27\%}
\newcommand{\dAearliest}{15.0\%}
\newcommand{\dAdelayup}{+12.6\%}
\newcommand{\dkhfrozen}{0.05}
\newcommand{\dkhearly}{0.07}
\newcommand{\dAfmlist}{-0.4, -0.2, +0.0, +0.6, +1.4\%}
\newcommand{\dAfmlo}{-0.4\%}
\newcommand{\dAfmhi}{+1.4\%}
\newcommand{\fnotfmlo}{37.2\%}
\newcommand{\fnotfmhi}{40.0\%}
\newcommand{\fnotfmlist}{40.0, 39.6, 39.1, 38.3, 37.2\%}

\begin{document}

\title{Gravitational-wave background from supermassive black holes
without merger trees: Tidally regulated mergers and direct inference from the NANOGrav 15 yr data}

\author{Shin'ichiro Ando}
\affiliation{GRAPPA Institute, University of Amsterdam,
1098 XH Amsterdam, The Netherlands}
\affiliation{Kavli Institute for the Physics and Mathematics of the
Universe (Kavli IPMU, WPI), University of Tokyo, Kashiwa, Chiba 277-8583,
Japan}

\date{\today}

\begin{abstract}
We present a semi-analytic framework that computes the gravitational-wave
background (GWB) from supermassive black hole (SMBH) binaries by direct
integration of an $N$-body-recalibrated extended Press--Schechter
accretion kernel, bypassing Monte Carlo merger trees entirely.  The
calculation is resolution-free and fast enough to be
sampled directly against pulsar-timing data with no emulator or
precomputed model bank.
Between halo accretion and black-hole binary formation, we couple the
satellite's simulation-calibrated tidal mass loss continuously to
Chandrasekhar dynamical friction.  Satellites below a sharp, nearly
redshift-independent mass-ratio threshold, $\xi \simeq \xiperm$, are
stripped faster than they sink and stall: their drag collapses onto the
stellar core, and at $z=0$, \fstallperm\ of accreted satellites still
owe more than a Hubble time of sinking.  Counting also those still in
transit, \fstallpure\ are unmerged by $z=0$.  This lowers the GWB
amplitude by \dAdelayabs\ relative to fixed-mass merger-time formulas, and
gives a physical origin for wandering SMBHs whose predicted mass
budget matches cosmological simulations, although their number in a
Milky-Way-mass halo does not.  Combining this population model with
per-binary stellar-hardening turnovers and eccentric spectra, we sample
the five-parameter posterior directly against the NANOGrav 15 yr
Hellings--Downs-correlated free spectrum, \chg{excluding the
$15.8$\,nHz bin, whose excess power a subsequent chromatic-noise
reanalysis attributes to pulsar noise.  The data prefer a black-hole
mass normalization $\dkhval$\,dex above the locally calibrated
black-hole--bulge mass relation ($\dkhaspub$\,dex if that bin is
retained), consistent with earlier indications that the nanohertz
background is loud relative to local SMBH calibrations and robust to
the delay model, merger kernel, local mass calibration, and its
redshift evolution.  The spectrum is compatible with, but does not
require, eccentric binaries or dense stellar environments.}
\end{abstract}

\maketitle

\section{Introduction}
\label{sec:intro}

Pulsar timing arrays (PTAs) have reported evidence for a stochastic
gravitational-wave background (GWB) at nanohertz (nHz) frequencies,
with the Hellings--Downs~\cite{Hellings:1983fr} angular correlation
expected of gravitational waves: the NANOGrav 15 yr data set~\cite{NANOGrav:2023gor}, the EPTA/InPTA second data release~\cite{EPTA:2023fyk}, the Parkes PTA~\cite{Reardon:2023gzh}, the
Chinese PTA~\cite{Xu:2023wog}, and the MeerKAT PTA~\cite{MPTA:2025dr,MPTA:2025gwb}.
Writing the characteristic strain of a fiducial
gravitational-wave-driven binary population as
$\hc(f) = \Ayr\, (f/f_{\rm yr})^{-2/3}$ with
$f_{\rm yr} = 1\,{\rm yr}^{-1}$, the measured amplitude is
\chg{$\Ayr = \AyrCNM$~\cite{Agarwal:2026dlv}, a $12\%$ downward
revision of the originally reported
$\AyrDMX$~\cite{NANOGrav:2023gor} brought by customized chromatic
noise models~\cite{Larsen:2026pru}.}  In terms of the
timing-residual spectral index $\gamma$, defined by
$\hc \propto f^{(3-\gamma)/2}$, a free power-law fit gives
\chg{$\gamma = \gammaCNM$, shallower than the
gravitational-wave-driven $\gamma = 13/3$ but consistent with it
inside the $95\%$ region~\cite{Agarwal:2026dlv}.  The original
analysis gave the shallower $\gamma = \gammaDMX$, and much of the
motivation for invoking spectral modification rested on
it~\cite{NANOGrav:2023gor}.}  The leading astrophysical
interpretation is the superposition of inspiraling supermassive black
hole (SMBH) binaries formed in galaxy mergers, anticipated since the
earliest PTA forecasts~\cite{Rajagopal:1994zj,Jaffe:2002rt,Wyithe:2002ep,Sesana:2008mz,
NANOGrav:2023hfp}, although early-Universe sources remain
statistically viable interpretations of the same data~\cite{NANOGrav:2023hvm}.  Extracting black-hole and
galaxy-evolution physics from this signal requires a population model
connecting halo mergers, through galaxy pairing and binary formation,
to the sub-pc hardening where nHz waves are emitted.

Most existing predictions build this chain on merger trees: Monte Carlo
realizations of extended Press--Schechter (EPS) branching~\cite{Press:1973iz,Bond:1990iw,Lacey:1993iv}, halo catalogs from
$N$-body simulations, or hydrodynamic volumes~\cite{Sesana:2008mz,Sesana:2012ak,IzquierdoVillalba:2021prf,
Kelley:2017lek}.  Trees are flexible but carry two structural costs.
First, a mass resolution must be chosen, and the shot noise of discrete
branchings propagates into the predicted spectrum.  Pushing the
resolution down to the progenitors that matter for seed-black-hole
populations is expensive.  Second, and more consequentially for
inference, a tree-based pipeline is too slow to be called inside a
sampler.  Posterior exploration must instead go through precomputed
model banks or emulators.  The NANOGrav astrophysical interpretation~\cite{NANOGrav:2023hfp} samples a Gaussian-process emulator trained on
a \holodeck\ library.
At the opposite extreme, fully analytic
EPS-based calculations~\cite{Ellis:2023dgf,Ellis:2023owy} evaluate the
GWB as a direct integral over the Lacey--Cole merger
kernel~\rev{\cite{Lacey:1993iv} (hereafter LC93)}.  This
route is fast and resolution-free, but it is built on the \emph{raw}
EPS rate, whose known $\mathcal{O}(1)$ inaccuracies against $N$-body
merger rates~\cite{Fakhouri:2010id} then propagate into the signal,
and it reduces the halo-merger-to-binary delay to a constant merger
probability.

This paper develops a middle path with three ingredients.

\emph{(i) A tree-free, $N$-body-calibrated merger integral.}  We use
the subhalo accretion-rate distribution of Ref.~\cite{Yang:2011rf}
(hereafter Y11), the EPS branching statistics recalibrated against
simulations, as implemented in the semi-analytic subhalo code
\sashimi\footnote{\url{https://github.com/shinichiroando/sashimi-c}}~\cite{Hiroshima:2018kfv,Ando:2019xlm,Dekker:2021bos,Ando:2024kpk}, and treat it as
a continuous merger kernel.  Multiplied by the Sheth--Tormen mass
function~\cite{Sheth:1999mn} and integrated over host mass, accretion
redshift, and satellite mass, this kernel counts every branch--branch
crossing of the ensemble of merger trees exactly once, with no
realization noise and no resolution floor
(Sec.~\ref{sec:framework}).
Against the
Fakhouri--Ma--Boylan-Kolchin (F10) $N$-body rate~\cite{Fakhouri:2010id}, the calibrated kernel reproduces the GWB
amplitude to \dAkernelabs, whereas the raw \rev{LC93} kernel used in
Refs.~\cite{Ellis:2023dgf,Ellis:2023owy} deviates by \dAlcabs\ (Sec.~\ref{sec:kernels}).  The calibration buys a factor of
\dAkernelgain\ in kernel systematics at zero runtime cost.

\emph{(ii) Merger delays regulated by subhalo tidal evolution.}
Between halo accretion and black-hole binary formation, the satellite
must sink to the halo center by dynamical friction while tides strip
its mass.  The two compete, because the friction force scales with the
satellite's surviving bound mass.  Almost all PTA population models
evaluate this stage at \emph{fixed} (infall) satellite mass~\cite{Kitzbichler:2008cy,BoylanKolchin:2007ku,Sesana:2012ak,
Chen:2018znx,NANOGrav:2023hfp}, or impose no delay at all~\cite{Ellis:2023dgf,Simon:2023khd}.  Reference~\cite{IzquierdoVillalba:2021prf} lets stripping set the \emph{starting
radius} of a subsequent fixed-mass phase, but the sinking mass is
frozen once that handoff occurs.  \rev{Integrating the orbit while the
satellite is stripped is, by contrast, routine in satellite dynamics, where
the two are solved together against simulation-calibrated
recipes~\cite{Jiang:2020rdj,Boldrini:2026thz}.  We bring that treatment into the
nHz-background calculation, coupling \sashimi's stripping
law~\cite{Hiroshima:2018kfv} \emph{continuously} to Chandrasekhar dynamical
friction~\cite{Chandrasekhar:1943ys} throughout the orbital decay
(Sec.~\ref{sec:delays}), and ask what it changes for the background and for
the black holes left behind.}

The competition then acquires a sharp, nearly redshift-independent
critical mass ratio.  Satellites with $\xi \lesssim \xiperm$ (where $\xi$ is the satellite-to-host mass ratio) are
stripped faster than they sink, their drag collapses onto the surviving
stellar core, and the orbital decay all but halts: at $z=0$,
\fstallperm\ of accreted satellites still owe more than a Hubble time of
sinking.  A fixed-mass timescale leaves such satellites behind too, but
selects them by epoch as much as by mass ratio, and the two therefore
disagree about \emph{which} black holes are left wandering
(Sec.~\ref{sec:stall}).
The amplitude effect is a reduction of \dAdelayabs, well inside the
current amplitude uncertainty, which is why fixed-mass formulas have
served PTA predictions adequately.  The
prize lies elsewhere: the same calibrated dynamics predict, with no
extra freedom, the black holes that have not arrived at their host's
center by $z=0$\rev{, a population
tracked in cosmological simulations and in semi-analytic
models~\cite{Tremmel:2018wsb,Ricarte:2021qlk,IzquierdoVillalba:2020hfk,Untzaga:2024gvk}}.  Their wandering mass fraction
in galaxy-scale halos is \fwandgal\ under the tidally regulated delay,
matching without tuning the $\sim$$10\%$ seen in cosmological
hydrodynamic simulations, whereas the fixed-mass delay strands only
\fwandgalbk\ (Sec.~\ref{sec:res_stall}).
This census is
\emph{parameter-free} and exactly independent of the fitted mass
normalization, which makes it an electromagnetic test of the delay
physics rather than a consequence of fitting the background.  The
independence matters, because the fit alone barely tests the delay: the
sampled mass normalization absorbs the \dAdelayabs\ offset, and the two
delay models give nearly indistinguishable spectra
(Sec.~\ref{sec:results}).  Earlier work on the GWB imprint of stalling
imposed the stalled fraction externally~\cite{Dvorkin:2017vvm}.  Here, it
and its mass-ratio boundary are both outputs of the dynamics.

\rev{\emph{(iii) The whole chain inside the sampler.}}  Because the integral framework
is deterministic and nearly all of it is independent of the black-hole
parameters, a full likelihood evaluation (population, delays,
per-binary environmental spectral turnover, and eccentric harmonics)
costs milliseconds after caching (Sec.~\ref{sec:priors}).
We therefore sample the five-parameter
posterior directly against the NANOGrav 15 yr Hellings--Downs
free-spectrum data~\cite{NANOGrav:2023gor,Lamb:2023jls}, calling the
exact model inside the Markov chain Monte Carlo (MCMC) sampler with no
emulator, no interpolation in parameter space, and no model bank
(Sec.~\ref{sec:inference}).  \rev{Emulator-free inference is itself established, and the way it is
achieved is instructive.  One route parametrizes the merger rate directly
from observed galaxy statistics, bypassing halos
entirely~\cite{Middleton:2015oda,Middleton:2020asl,Chen:2018znx}.  The
other keeps halos but takes the raw analytic EPS rate and collapses the
whole halo-merger-to-coalescence stage into a single constant
efficiency~\cite{Ellis:2023dgf,Raidal:2024odr}.  A third parametrizes an
EPS black-hole population and confronts pulsar-timing with
high-redshift data in one Bayesian fit, again carrying the delay as a
fitted parameter~\cite{Toubiana:2024bil}.  Each buys its speed by
simplifying the stage this paper is about.  The calculation here keeps the
halo kernel, calibrated against $N$-body rates, and integrates the delay
from the stripping and sinking equations, while staying cheap enough to be
the sampled model itself.}
The primary result of this direct inference is
a black-hole mass normalization $\dkhval$\,dex above the locally
calibrated relation, \chg{preferred by the gravitational-wave amplitude at
\pdkhT\ of the posterior and robust to the delay, kernel, and
calibration systematics (Sec.~\ref{sec:results}).  The fit excludes the
one frequency bin whose excess power a later chromatic-noise reanalysis
attributes to pulsar noise (Sec.~\ref{sec:likelihood}); retaining it
would give $\dkhaspub$\,dex.}

The paper is organized as follows.  Section~\ref{sec:framework}
presents the tree-free merger integral, the galaxy--black-hole
painting, and the kernel comparison.  \rev{Section~\ref{sec:delays}
follows a pair from halo accretion to coalescence, through the tidally
regulated sinking stage, its fixed-mass counterpart, and the residual
hardening delay, and identifies the population that never arrives.
Section~\ref{sec:spectrum} turns the surviving population into an
observed spectrum, with the loss-cone turnover and eccentricity.}
Section~\ref{sec:inference} defines the likelihood and priors, and
Sec.~\ref{sec:results} presents posteriors under both the fixed-mass
and tidal delay models, together with the delay model's parameter-free
stalled fractions and their predictions, falsifiable with
wandering-black-hole demographics.
We discuss systematics, the relation to previous work, and the
extension to the LISA band in Sec.~\ref{sec:discussion}, and conclude
in Sec.~\ref{sec:conclusions}.  Throughout we adopt the Planck 2018
cosmology~\cite{Planck:2018vyg} used by \sashimi.

\section{Merger rates without merger trees}
\label{sec:framework}

\subsection{The master integral}
\label{sec:master}

A Monte Carlo merger tree encodes two distinct pieces of information:
the assembly history of individual halos, and the statistical rate at
which branches of given masses join.  For a population-level observable
such as the GWB, only the latter is needed, and EPS theory provides it
in closed form.  Our starting point is the subhalo accretion-rate
distribution of Y11 (their Model III), as
implemented and used to anchor the evolved subhalo population in
\sashimi~\cite{Hiroshima:2018kfv}: the mean number of subhalos of mass
$m_a$ accreted per logarithmic mass and per logarithmic $(1+z_a)$
interval onto a host that reaches mass $M_0$ at $z=0$,
\begin{equation}
  \mathcal{N}_a\!\left(m_a, z_a \,\middle|\, M_0\right)
  \equiv \frac{d^2 N}{d\ln m_a\, d\ln(1+z_a)} .
  \label{eq:Na}
\end{equation}
This function is constructed from the same excursion-set branching
statistics as a Monte Carlo tree, but with its normalization and shape
recalibrated against cosmological simulations~\cite{Yang:2011rf}.  It is
continuous in both arguments, and therefore carries no resolution floor
and no realization scatter.

Every merger in the ensemble of trees is an accretion event onto some
main branch.  We therefore weight Eq.~\eqref{eq:Na} by the comoving
abundance of main branches at each epoch.  Writing $M(z_a|M_0)$ for the
mean main-branch mass history~\cite{Correa:2015dva} and changing
variables from the $z=0$ mass $M_0$ to the mass at accretion, the
comoving density of halo--halo mergers per unit host mass, accretion
redshift, and satellite mass is
\begin{equation}
  \frac{d^3 n}{d\ln M_0\, d\ln(1+z_a)\, d\ln m_a}
  = \frac{dn}{d\ln M}
    \frac{\partial \ln M}{\partial \ln M_0}\,
    \mathcal{N}_a ,
  \label{eq:master}
\end{equation}
where $dn/d\ln M$ is the Sheth--Tormen mass function~\cite{Sheth:1999mn} evaluated at the main-branch mass $M(z_a)$ and
epoch $z_a$, applied at its native virial mass definition
(Appendix~\ref{app:valid}), and using the same $\sigma(M)$ as
\sashimi~\cite{Ludlow:2016ifl}.
The choice of mass-function fit moves the amplitude by between
$\dAtinkfit$ and $\dAsiafit$, quantified in Sec.~\ref{sec:systematics}.  
Three approximations are inherited: the mean
mass history replaces the distribution of histories, correlations
between successive accretions within one tree are neglected
(irrelevant for one-point statistics such as $\hc$), and
satellite--satellite mergers inside a common host are not counted.
Each pair is counted once: the Y11 kernel is defined up to
its natural boundary $m_a \le M_0/2$, and the residual sensitivity to
how the major-merger edge is treated is contained in the end-to-end
kernel comparison of Sec.~\ref{sec:kernels}.

The triple integral of Eq.~\eqref{eq:master} is evaluated on a
$36\times 48\times 64$ logarithmic grid in $(M_0, 1+z_a, m_a)$ spanning
$M_0 = 10^{11}$--$10^{15.7}\,\Msun$, $z_a = 0.02$--$6$, and
$m_a = 10^{10}$--$10^{15.5}\,\Msun$.  Doubling every grid dimension
changes the amplitude by $0.1\%$.

\subsection{Galaxies and black holes}
\label{sec:painting}

Each side of a halo--halo merger is painted with a galaxy through the
stellar-mass--halo-mass relation of Ref.~\cite{Moster:2012fv}.  The
satellite is painted at its infall mass $m_a$.  Once it enters the host,
it is quenched and tidally stripped, and its black hole is held fixed
thereafter.  The host keeps accreting and merging until the galaxies
coalesce.  We therefore advance its main-branch mass from $M(z_a)$ to
its value at the galaxy-merger epoch (Sec.~\ref{sec:delays}) before
assigning its black hole, a co-evolution consistent with the delay
physics, and hold the primary fixed through the subsequent binary
hardening.
Explicitly,
\begin{equation}
  \frac{M_\star}{M} = 2N
  \left[\left(\frac{M}{M_1}\right)^{-\beta}
      + \left(\frac{M}{M_1}\right)^{\gamma_M}\right]^{-1},
  \label{eq:moster}
\end{equation}
with the best-fit parameters:
$\log_{10}(M_1/\Msun) = 11.590 + 1.195\, z/(1+z)$,
$N = 0.0351 - 0.0247\, z/(1+z)$,
$\beta = 1.376 - 0.826\, z/(1+z)$, and
$\gamma_M = 0.608 + 0.329\, z/(1+z)$~\cite{Moster:2012fv}.  Stellar bulges are assigned black holes
through the Kormendy--Ho relation~\cite{Kormendy:2013dxa},
\begin{equation}
  \Mbh = 10^{\dkh}\, 0.49\times 10^{9}\,\Msun\,
         \left(\frac{M_{\star}}{10^{11}\,\Msun}\right)^{1.16},
  \label{eq:kh13}
\end{equation}
where we take $M_{\rm bulge} \simeq M_{\star}$ appropriate for the
massive systems that dominate the signal, and $\dkh$ parameterizes a
log-normalization offset.  Each black-hole mass carries an independent lognormal
scatter of width $\sdex$ at fixed halo, which enters the signal through
the average $\langle \mathcal{M}_c^{5/3} \rangle$ of the chirp mass
$\mathcal{M}_c = (m_1 m_2)^{3/5}/(m_1+m_2)^{1/5}$.  We evaluate this
average with a $5\times5$ Gauss--Hermite quadrature.  Both galaxies
must exceed $M_\star > 10^{8}\,\Msun$ for the pair to count.

Because every massive galaxy is observed to host a central black hole
(an assumption already embedded in the calibration of
Eq.~\eqref{eq:kh13}), we set the occupation fraction to unity.  Two
clarifications guard against misreading this choice.  First, any
mass-dependent \emph{shape} of the occupation function (a low-mass
cutoff scale $M_{\rm occ}$ reflecting seeding physics) is invisible in
the PTA band: the halos that dominate $\hc$ have
$M \gtrsim 10^{12}\,\Msun$, orders of magnitude above any plausible
occupation cutoff.  The PTA signal therefore cannot constrain seeding,
and that information lives in the LISA band
(Sec.~\ref{sec:discussion}).
Second, a global occupation deficit $f_{\rm occ} < 1$ at the massive
end would enter the signal only through the combination
$f_{\rm occ}^2 \langle \mathcal{M}_c^{5/3} \rangle$, exactly
degenerate with a shift of $\dkh$.  Our $\dkh$ posterior should be
read as constraining this combination.

\subsection{Kernel systematics: what the calibration buys}
\label{sec:kernels}

\begin{figure*}
\includegraphics[width=\textwidth]{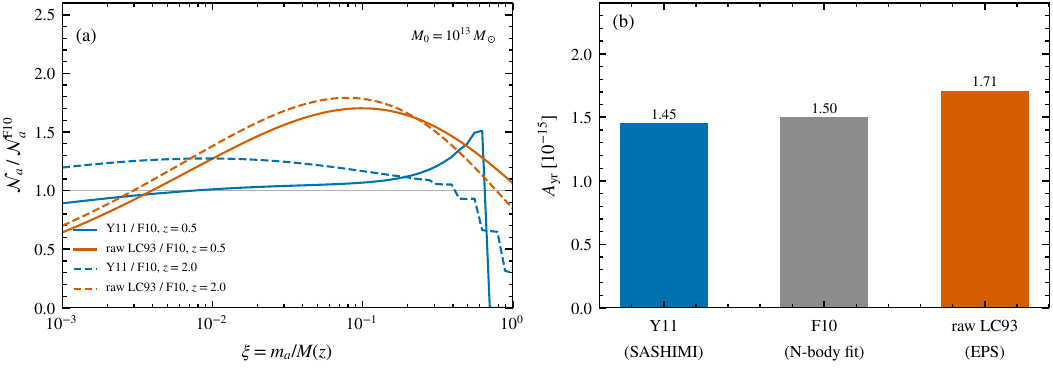}
\caption{(a) The three halo merger kernels $\mathcal N_a$ [Eq.~\eqref{eq:Na}] 
divided by that of the F10 $N$-body rate~\cite{Fakhouri:2010id},
as a function of mass ratio $\xi$, for a host with
$M_0 = 10^{13}\,\Msun$ at $z=0.5$ (solid) and $z=2$ (dashed): the
$N$-body-recalibrated Y11 kernel~\cite{Yang:2011rf} used here
(blue) and the raw \rev{LC93} EPS kernel~\cite{Lacey:1993iv} used by
Refs.~\cite{Ellis:2023dgf,Ellis:2023owy} (vermilion).  (b) The
resulting GWB amplitude $\Ayr$ at the fiducial configuration (tidal
delay, $\thard = 0.5$\,Gyr, $\sdex = 0.33$), with only the kernel
changed.  }
\label{fig:kernels}
\end{figure*}

The EPS merger rate has a known conceptual ambiguity (it is
asymmetric under exchanging progenitors) and $\mathcal{O}(1)$
normalization offsets against simulations.  To quantify what the
Y11 recalibration buys, we repeat the entire calculation
(Fig.~\ref{fig:kernels}) with the kernel $\mathcal N_a$ in Eq.~\eqref{eq:master}
swapped for (i) the raw \rev{LC93} rate~\cite{Lacey:1993iv} [their
Eq.~(2.18), reduced to the per-host form in Appendix~\ref{app:lc}],
the kernel underlying the tree-free calculations of
Refs.~\cite{Ellis:2023dgf,Ellis:2023owy}, and (ii) the
F10 fit to the Millennium simulations~\cite{Fakhouri:2010id}, which we treat as the $N$-body reference.
Figure~\ref{fig:kernels} shows the rates at a representative host mass
and the GWB amplitudes obtained with each kernel under otherwise
identical assumptions.  The Y11 kernel tracks the $N$-body
reference to within $\pm(20$--$50)\%$ pointwise over the mass ratios
and redshifts that matter, and to \dAkernelabs\ in the amplitude (the
pointwise residuals average out in the triple integral), while the raw
EPS kernel is biased high by \dAlcabs\ in amplitude.  Since
$\hc \propto \sqrt{{\rm rate}}$ at fixed masses, kernel systematics
at this level are subdominant to the delay and black-hole-mass
uncertainties that we treat as free parameters below.

We use F10 as the $N$-body reference but not as the fiducial kernel,
for two reasons.  The first is internal consistency: Y11 is the
accretion distribution on which \sashimi's evolved subhalo population
is built, and the tidal stripping law that regulates our delays
(Sec.~\ref{sec:tidal}) is calibrated within that same framework.
Pairing F10 rates with \sashimi\ stripping would splice together two
independent calibrations.  The second is portability: F10 is a fitting
formula tied to the Millennium simulations and their cosmology, valid
over the mass ratios and redshifts sampled there, whereas Y11 is
analytic in $\sigma(M)$ and therefore adopts our cosmology
self-consistently and extends smoothly to the low masses and high
redshifts of the seed regime, with no resolution floor.  The
\dAkernelabs\ amplitude agreement between the two shows that this choice is not a
limiting systematic in any case.  The kernel counts accretion onto the
main branch.  Satellite--satellite sinking, which $N$-body measurements
find can rival the central--satellite channel at low mass
thresholds~\cite{Jiang:2026hqh}, is not included, and alternative
routes to the same rate are set out in
Refs.~\cite{Leonova:2026xhz,Capelo:2026gij}.

\section{\rev{Merger delays: from halo accretion to coalescence}}
\label{sec:delays}

\subsection{Stripping coupled to sinking}
\label{sec:tidal}

\sashimi's evolved subhalo population is built on a tidal stripping
law calibrated against cosmological simulations~\cite{Hiroshima:2018kfv},
\begin{equation}
  \frac{dm}{dt}
  = -A(M,z)\, \frac{m}{\tau_{\rm dyn}}
    \left(\frac{m}{M}\right)^{\zeta(M,z)},
  \label{eq:strip}
\end{equation}
with $A \sim \mathcal{O}(1)$ and $\zeta \lesssim 0.02$ weak functions
of host mass and redshift.  We couple Eq.~\eqref{eq:strip} to the orbital
decay of the same object under Chandrasekhar dynamical friction~\cite{Chandrasekhar:1943ys,Binney:2008} in a singular isothermal host
with circular velocity $V_c$:
\begin{align}
  r\,\frac{dr}{dt}
  &= -0.428\, \ln\Lambda\, \frac{G\, m_{\rm eff}(t)}{V_c},
  \nonumber\\
  \ln\Lambda &= \ln\!\left(1 + M/m_{\rm eff}\right).
  \label{eq:sink}
\end{align}
The drag is proportional to the satellite's \emph{surviving} bound
mass.  We take $m_{\rm eff}(t) = m(t) + m_\star$: the dark matter
remnant of Eq.~\eqref{eq:strip} plus the satellite's stellar mass,
which acts as a floor because the tightly bound stellar core (and its
black hole) survives the stripping that removes the outer halo.
\rev{Because $m_{\rm eff} \to m_\star > 0$, the drag never switches off:
Eq.~\eqref{eq:sink} keeps draining $r^2$ at a fixed rate, and every
satellite reaches the center eventually.  Throughout this paper a
``stalled'' satellite is therefore one whose \emph{remaining} sinking time
at $z=0$ exceeds a Hubble time, not one frozen forever.}
The competition between stripping and sinking is decided by how much
orbital decay the satellite completes before its dark matter envelope
is gone.  Treating the stellar core as unstrippable maximizes the
residual drag, and the stalled fractions below are therefore lower
bounds.

Equations~\eqref{eq:strip} and \eqref{eq:sink} integrate as a single
vectorized march in redshift over the entire $(M_0, z_a, m_a)$ grid,
with no per-orbit ODE solves.  The march costs a few seconds once per
population model and, being independent of the sampled parameters, adds
nothing to the cached per-likelihood cost (Sec.~\ref{sec:priors}).  Its
numerical construction and convergence tests are collected in
Appendix~\ref{app:valid}.  A satellite is deemed merged with the
central galaxy when $r < f_{\rm merge}\, r_{\rm vir}(z)$ with
$f_{\rm merge} = 0.05$.  This threshold marks the assembly of the
\emph{galaxy} pair, not the onset of gravitational-wave emission,
which occurs much later at sub-pc separations
(Sec.~\ref{sec:spectrum}).  The result is insensitive to the
threshold: at the canonical $f_{\rm merge} = 0.05$ the non-coalesced
fraction is \fnotfidT, and varying $f_{\rm merge}$ over $0.02$--$0.10$
moves it between \fnotfmlo\ and \fnotfmhi\ while shifting the GWB
amplitude by only $\dAfmlo$ to $\dAfmhi$ (Appendix~\ref{app:valid}).

\subsection{The fixed-mass baseline}
\label{sec:bk08}

The delay between halo accretion and the formation of a bound black-hole
binary at the galaxy center is conventionally described by a
dynamical-friction merger timescale evaluated at infall.  The standard
choice is the $N$-body-calibrated formula of
Ref.~\cite{BoylanKolchin:2007ku} (hereafter BK08),
\begin{equation}
  \frac{\tau_{\rm df}}{\tau_{\rm dyn}}
  = 0.216\, \frac{(M/m_a)^{1.3}}{\ln\!\left(1+M/m_a\right)}\,
    e^{1.9\eta},
  ~~
  \tau_{\rm dyn} = \sqrt{\frac{r_{\rm vir}^3}{G M}},
  \label{eq:bk08}
\end{equation}
with $\eta$ the orbital circularity (we use the population mean
$\eta = 0.5$~\cite{Zentner:2004dq}) and $M \equiv M(z_a)$ the
main-branch host mass at the accretion epoch.  \rev{It is the comparison
model against which the tidal delay is judged.}

Equation~\eqref{eq:bk08} is a fit to simulations of live satellites
sinking in live hosts, and tidal mass loss is folded into its
coefficients \emph{on average}.  But the satellite mass appearing in it
is frozen at infall, and the formula therefore assigns a smooth, finite
merger time to \emph{every} mass ratio.  \rev{That time diverges as
$\xi^{-1.3}$, and the fixed-mass model too leaves a tail of satellites
unmerged by $z=0$, some owing more than a further Hubble time.  With the
mass frozen, however, the delay carries no threshold: $\tau_{\rm df}$
varies smoothly with $\xi$ and, through $\tau_{\rm dyn}(z_a)$, with the
epoch, and the boundary of its stalled set therefore runs diagonally in
$(\xi, z_a)$ rather than vertically (Sec.~\ref{sec:stall}).}  The tidal
coupling of Sec.~\ref{sec:tidal} instead
lengthens the sinking time \emph{selectively}, for the low-mass-ratio
satellites whose dark matter is stripped before they arrive.

\subsection{\rev{The residual hardening delay}}
\label{sec:thard}
\rev{

Reaching the halo center assembles the \emph{galaxy} pair.  The two black
holes must still travel from kpc separations to the sub-pc
orbits where nHz waves are emitted, by dynamical friction inside the
merged remnant~\cite{Dosopoulou:2016hbg} and through the final-parsec
regime~\cite{Begelman:1980vb,Milosavljevic:2002ht}.  We lump that journey
into a single residual delay $\thard$, common to the population and
sampled in the inference (Table~\ref{tab:params}).  A pair accreted at
$z_a$ therefore coalesces at the redshift $z_m$ reached by advancing $z_a$
first by the sinking time of Sec.~\ref{sec:tidal}, then by $\thard$.

A single timescale does not exhaust this stage.  The same hardening phase
also reshapes the emitted spectrum, through the time each binary spends
per unit frequency, and that aspect is modeled separately in
Sec.~\ref{sec:hardening}.  The two are complementary descriptions of one
stage rather than a double count: $\thard$ moves the coalescence
\emph{epoch}, and hence which pairs arrive by $z=0$, while the loss-cone
model moves \emph{where in frequency} the emitted power lands.}

\begin{figure}
\includegraphics[width=\columnwidth]{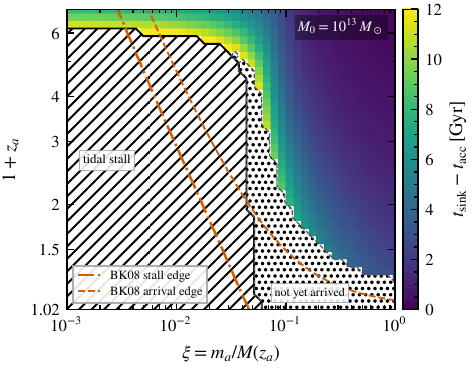}
\caption{Time from halo accretion to galaxy merger under the tidally
regulated delay, for satellites accreted by a host with
$M_0 = 10^{13}\,\Msun$ at redshift $z_a$ with mass ratio $\xi$.
\rev{This is a single host mass: the fractions quoted in the text are
sums over the whole host population and are smaller than the areas here
suggest, and Table~\ref{tab:census} lists both.}  Two
physically distinct regions fail to deliver a black hole to the nucleus
by $z=0$.  Hatched: the tidal stall proper, where stripping has
collapsed the drag onto the stellar core and the sinking time still
owed at $z=0$ exceeds a Hubble time.  Its boundary is a sharp, nearly
redshift-independent mass-ratio threshold at $\xi \simeq \xiperm$.
Dotted: satellites accreted too recently to complete a normal few-Gyr
infall before $z=0$.  These are still sinking with their drag intact
and would arrive within another Hubble time.
\rev{Vermilion: the corresponding boundaries under the fixed-mass BK08
delay, Eq.~\eqref{eq:bk08}, for the stall (dash-dotted) and for arrival by
$z=0$ (dashed), the region between them being its own not-yet-arrived
population.  Both sweep with accretion epoch instead of standing as a wall
in mass ratio.}}
\label{fig:stall}
\end{figure}

\subsection{The non-merging population}
\label{sec:stall}

Figure~\ref{fig:stall} shows the delay surface, and it separates two
physically distinct ways a satellite can fail to deliver its black hole
by $z=0$.

\rev{Satellites below a critical mass ratio effectively stop sinking.}
Stripping
removes their dark matter envelope, $m_{\rm eff}$ collapses onto the
stellar floor, and the residual drag cannot complete the decay: the
sinking time still owed at $z=0$ exceeds a Hubble time, and over the
inner part of the region it exceeds many Hubble times.  \rev{This is the
tidal stall.}  Because the competition between stripping and
sinking is set by the mass ratio and only weakly by the epoch, its
boundary is a sharp, nearly redshift-independent threshold at
$\xi \simeq \xiperm$ (hatched region of Fig.~\ref{fig:stall}).
\rev{Summed over the whole host population (weighted by merger rate and
occupation, and \emph{not} read off the $10^{13}\,\Msun$ panel, where the
fractions are roughly twice as large), \fstallperm\ of accreted
satellites are stalled in this sense.}

Satellites accreted very recently have simply not had time to finish a
normal, few-Gyr infall before $z=0$, whatever their mass ratio (dotted
region).  This is not a stall.  Their drag is intact, they are still
sinking, and they would arrive within another Hubble time.  A
population like this exists in any delay model, the fixed-mass baseline
included.  It adds a further \fstalltrans\ of accreted satellites, on the same
population-wide weighting, and it is what makes the non-merging region
extend to large $\xi$ as $z_a \to 0$.

Together the two give the \fstallpure\ of accreted satellites that have
not delivered a black hole to the nucleus by $z=0$, which is the
fraction the background actually sees.
Neither fraction depends on any of the five sampled parameters
(Table~\ref{tab:params}).  Both are parameter-free predictions of the
calibrated dynamics rather than outputs of the fit.

\rev{The fixed-mass baseline strands satellites as well: applying the same
criterion to Eq.~\eqref{eq:bk08} leaves \fstallbk\ of accreted satellites
owing more than a Hubble time at $z=0$.  The two boundaries differ in
shape.  Stripping and sinking both scale with the host dynamical time, and
their nearly epoch-independent ratio makes the tidal boundary a wall in
mass ratio alone: fitting $\xi_{\rm crit} \propto (1+z_a)^{s}$ gives
$s = \tiltwall$, against $s = \tiltbk$ for the fixed-mass delay, whose
$\tau_{\rm df}$ and available time both move with $z_a$ (dash-dotted curve
in Fig.~\ref{fig:stall}).  The two therefore select different satellites,
which a wandering-black-hole census can test
(Sec.~\ref{sec:res_stall}).}

Adding the residual hardening delay $\thard$ of \rev{Sec.~\ref{sec:thard}},
the fraction of pairs that have not coalesced
by $z=0$ is \fnotfidT\ at the fiducial $\thard = 0.5$\,Gyr, against
\fnotfidB\ under the BK08 baseline.  The GWB amplitude drops by
\dAdelayabs, from $\Ayr = \Ayrbk$ to $\Ayrtidal$ at fiducial
parameters.  A shift of that size sits well inside the current
amplitude uncertainty, which explains why fixed-mass formulas have
served PTA predictions adequately.  In the full inference, it is
absorbed almost entirely by $\dkh$, and the fitted spectrum barely
distinguishes the two delay models (Sec.~\ref{sec:results}).  \rev{What
does distinguish them is \emph{which} satellites are left behind, a
deterministic function of $(M_0, z_a, \xi)$ once the calibration is
fixed.}
The model turns it into falsifiable
predictions for kpc-scale stalled companions and wandering black holes
with definite mass and host-mass distributions~\cite{Tremmel:2018wsb,Ricarte:2021qlk,Barausse:2020mdt}, rather than
absorbing the non-merging fraction into an opaque efficiency parameter.

Radial orbits sink faster [the $e^{1.9\eta}$ of Eq.~\eqref{eq:bk08}],
but are also stripped faster.  Marginalizing both effects over the
infall circularity distribution~\cite{Zentner:2004dq} changes the
amplitude by $\le \dAorbit$, and we therefore use the mean orbit
throughout (Appendix~\ref{app:valid}).

\section{\rev{The emitted spectrum}}
\label{sec:spectrum}

\rev{Sections~\ref{sec:framework} and~\ref{sec:delays} deliver the
population of coalescing binaries.  We now assemble it into an observed
spectrum, adding the two per-binary effects that shape the nHz band:
the stellar-hardening turnover and orbital eccentricity.  Both carry a
sampled parameter, and we set them out in the detail that requires.}

\subsection{The gravitational-wave-driven baseline}
\label{sec:phinney}

\begin{figure}
\includegraphics[width=\columnwidth]{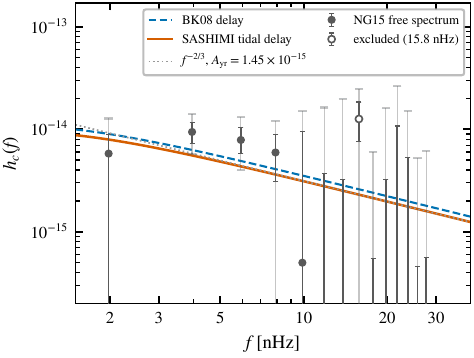}
\caption{Characteristic strain at fiducial parameters
($\dkh = 0$, $\sdex = 0.33$, $\thard = 0.5$\,Gyr, circular binaries,
fiducial nuclear density) under the two delay models, compared with
the NANOGrav 15 yr Hellings--Downs free-spectrum posteriors (points:
medians with 68\% and 95\% intervals).  \chg{The open point at
$15.8$\,nHz is plotted but excluded from the likelihood: a
chromatic-noise reanalysis of the same data attributes that excursion
to non-stationary transients in PSR~J1713+0747~\cite{Agarwal:2026dlv}
(Sec.~\ref{sec:likelihood}).}  The tidal delay removes
stalled satellites and lowers the amplitude by \dAdelayabs.}
\label{fig:spectrum}
\end{figure}

For circular, purely gravitational-wave-driven binaries, the
characteristic strain is given as~\cite{Phinney:2001di}
\begin{equation}
  \hc^2(f) = \frac{4 G^{5/3}}{3 \pi^{1/3} c^2}\, f^{-4/3}
  \int d^3 n \;
  \big\langle \mathcal{M}_c^{5/3} \big\rangle\,
  \frac{P_{\rm merge}}{(1+z_m)^{1/3}} ,
  \label{eq:phinney}
\end{equation}
where the integral runs over the merger density of
Eq.~\eqref{eq:master}, \rev{$z_m$ is the coalescence redshift built up in
Sec.~\ref{sec:thard}}, and $P_{\rm merge}$
selects pairs that coalesce before $z=0$ (for the tidal delay model,
stalled cells have $P_{\rm merge} = 0$).  Equation~\eqref{eq:phinney} yields
the familiar $f^{-2/3}$ spectrum ($\gamma = 13/3$).  At fiducial
parameters the amplitude is $\Ayr = \Ayrtidal$ (tidal
delay), within the population spread of published models and just
below the NANOGrav 15 yr measurement (Fig.~\ref{fig:spectrum}).

\subsection{Stellar hardening and the spectral turnover}
\label{sec:hardening}

\rev{This is the spectral face of the stage that Sec.~\ref{sec:thard}
lumped into $\thard$.}  At wide separations, a bound binary hardens not by radiating
gravitational waves but by ejecting stars: three-body slingshot
encounters with stars on low-angular-momentum orbits (the binary's
``loss cone'') extract orbital energy and shrink the binary
semimajor axis $a$ at the rate
$da/dt = -H G \rho\, a^2/\sigma$.  Here, $\rho$ is the
stellar mass density and $\sigma$ the stellar velocity dispersion of
the surrounding nucleus, and $H \simeq 16$ is the dimensionless
hardening rate measured in three-body scattering experiments~\cite{Quinlan:1996vp,Sesana:2015haa}.
Equating this rate to the Peters gravitational-wave hardening rate~\cite{Peters:1964zz} defines, per binary, a transition separation
and source-frame frequency
\begin{align}
  a_t^5 &= \frac{64}{5 H}\,
          \frac{G^2 m_1 m_2 (m_1+m_2)\, \sigma}{c^5 \rho},
  \nonumber\\
  f_t &= \frac{1}{\pi} \sqrt{\frac{G (m_1+m_2)}{a_t^3}} ,
  \label{eq:ft}
\end{align}
where $f_t$ is the gravitational-wave frequency (twice the orbital
frequency) at $a_t$.
Below $f_t$ a binary spends less time per logarithmic frequency than
gravitational waves alone would allow, suppressing its contribution
by $S(f) = [1+(f_t/f_r)^{10/3}]^{-1}$ with $f_r = (1+z_m) f$ the
source-frame frequency.  The prospect of constraining such
environmentally induced turnovers with PTA spectra was established
with parametrized models in
Refs.~\cite{Sampson:2015ada,Taylor:2016ftv}.  The
stellar environment is assigned per binary by inverting the
$\Mbh$--$\sigma$ relation~\cite{Kormendy:2013dxa} and taking a
singular isothermal density at the influence radius,
$\rho_{\rm inf} = \sigma^6 / (2\pi G^3 M_{\rm bin}^2)$, scaled by a
free factor $\rhob$ that absorbs the (substantial) uncertainty in
nuclear densities and in $H$.  Only the combination $H\rho$ enters
Eq.~\eqref{eq:ft}.  Gas-driven hardening in gas-rich nuclei~\cite{Kocsis:2010xa} is not
modeled separately.  To the extent that it accelerates the traversal
of this phase, it acts like a larger effective $\rhob$.  Stellar
hardening alone moves the spectral slope only modestly: even with
$\rhob = 10$ the effective slope in the 2--30 nHz band goes from
$\gamma_{\rm eff} \simeq 4.3$ to $4.1$.  \chg{Measured against the
originally reported $\gamma = \gammaDMX$ that was a clear shortfall.
Measured against the revised $\gamma = \gammaCNM$ it is a
$\sim$$1\sigma$ difference, and stellar hardening on its own is no
longer excluded.}

\subsection{Eccentricity}
\label{sec:ecc}

Eccentric binaries radiate in a comb of harmonics of the orbital
frequency~\cite{Peters:1963ux} and circularize as they shrink~\cite{Peters:1964zz}.  Both effects move power from low to high
frequencies and flatten the spectrum in the PTA band~\cite{Enoki:2006kj,Chen:2016zyo,Raidal:2024odr}.  We assign each binary an
eccentricity $\et$ at its own transition frequency $f_t$, the same
per-binary $f_t$ of Eq.~\eqref{eq:ft}, hold it constant through the
stellar-hardening phase, and evolve it along the Peters track above
$f_t$.  A single $\et$ is therefore shared by the population, but it is
imposed at each binary's own $f_t$ rather than at a common frequency,
and above $f_t$, it is not held fixed: the binary circularizes along the
Peters track.  $\et$ is thus the eccentricity at the end of the
stellar-hardening phase, where gravitational-wave emission takes over.
In reality, the transition eccentricity carries a distribution, set by
the three-body scattering history that pumps eccentricity during
hardening, and $\et$ should be read as an effective, population-typical
value rather than a universal one.  Modeling that spread would smooth
the harmonic structure of $\Phi$ below, and we do not attempt it here.
The population spectrum generalizes
Eq.~\eqref{eq:phinney} by one extra per-binary factor
$\Phi(f_r/f_t; \et)$ in the integrand, replacing $S(f)$, with
\begin{equation}
  \Phi(y; \et) = \sum_{n\ge1}
  \left(\frac{2}{n}\right)^{2/3}
  \frac{g(n, e_n)}{F(e_n) + x_n^{-10/3}},
  ~~
  x_n = \frac{2 y}{n},
  \label{eq:phi}
\end{equation}
where $g(n,e)$ is the Peters--Mathews harmonic power~\cite{Peters:1963ux}, $F(e) = \sum_n g(n,e)$ the enhancement factor,
$e_n = e(x_n)$ the eccentricity on the circularization track, and the
$x_n^{-10/3}$ term carries the stellar-hardening suppression of
Sec.~\ref{sec:hardening} [at $\et = 0$,
Eq.~\eqref{eq:phi} reduces exactly to $S(f)$].  Because $\Phi$ depends on
the cell only through $y = f_r/f_t$, we tabulate it once per
likelihood call and interpolate.  \rev{That the eccentric,
environmentally hardened spectrum of any binary collapses onto one
reference curve under a frequency rescaling is the observation of
Ref.~\cite{Chen:2016zyo}, who introduced it, in analytic fitted form, for
exactly this purpose.  Using it keeps the eccentric model as cheap as the
circular one, which is what allows $\et$ to be sampled rather than
scanned.  Reference~\cite{Raidal:2024odr} reaches the same end by a
different semi-analytic route.}
The harmonic truncation is chosen adaptively
to capture $99.9\%$ of $F(\et)$.

\begin{figure}
\includegraphics[width=\columnwidth]{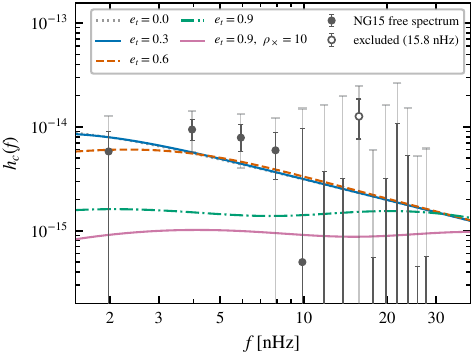}
\caption{Spectral shapes for the tidal-delay population at fiducial
parameters, varying the transition eccentricity $\et$ and (for one
curve) the nuclear density factor $\rhob$.  Eccentricity moves power
across the band: the slope flattens toward the measured
\chg{$\gamma = \gammaCNM$} while the amplitude near
$f = 1\,{\rm yr}^{-1}$ increases.  Data points as in
Fig.~\ref{fig:spectrum}.
}
\label{fig:ecc}
\end{figure}

Figure~\ref{fig:ecc} shows the model family: increasing $\et$
simultaneously flattens the PTA-band slope (reaching
$\gamma_{\rm eff} \approx 3.3$ at $\et \simeq 0.8$ and overshooting
to $\approx 3.0$ by $\et = 0.9$) and \emph{raises} the amplitude at
$f = 1\,{\rm yr}^{-1}$ by $\sim$10\%, both in the direction of the
NANOGrav data.

\section{Inference from the NANOGrav 15 yr free spectrum}
\label{sec:inference}

\subsection{Likelihood}
\label{sec:likelihood}

We use the NANOGrav 15 yr Hellings--Downs-correlated free-spectrum
posteriors in their official kernel-density-estimate (KDE)
representation~\cite{Lamb:2023jls,ng15_kde_zenodo}, the same data product that
underlies the {\sc ceffyl}/{\sc PTArcade} refitting frameworks~\cite{Lamb:2023jls,Mitridate:2023oar}.
The
free-spectrum analysis measures, at each frequency
$f_i = i/T_{\rm span}$ ($T_{\rm span} = 16.03$\,yr), the posterior of
the excess-power amplitude $\rho_i$, related to the strain by
\begin{equation}
  \rho_i^2 = \frac{\hc^2(f_i)}{12 \pi^2 f_i^3\, T_{\rm span}} .
  \label{eq:rho}
\end{equation}
In the factorized-likelihood approximation, the bins are treated as
independent (an approximation validated against the full
hierarchical PTA likelihood in Ref.~\cite{Lamb:2023jls}), so that
\begin{equation}
  \ln \mathcal{L}(\bm{\theta})
  = \sum_{i \in \mathcal{I}} \ln p_i\!\left(
      \log_{10} \rho_i\big[\hc(f_i;\bm{\theta})\big] \right),
  \label{eq:like}
\end{equation}
where $\bm{\theta}$ collects the five sampled population parameters
listed in Table~\ref{tab:params} and $p_i$ are the per-bin KDE
densities.  \chg{The index set $\mathcal{I}$ is the lowest 14 bins,
matching the frequency range of the NANOGrav Hellings--Downs
analyses~\cite{NANOGrav:2023gor}, less the eighth,
$f_8 = 15.8$\,nHz.}  Higher bins are white-noise dominated.

\chg{The exclusion is ours, and the reason needs stating precisely.
After the free spectrum used here was released,
Ref.~\cite{Agarwal:2026dlv} re-analyzed the same data set with
customized chromatic noise models~\cite{Larsen:2026pru} and reported
reduced correlated power in several low-frequency modes, the largest
reduction being at $15.8$\,nHz.  That bin had been flagged as an
excursion from power-law behavior, and they find the excursion to be a
noise-model-dependent effect, which they attribute to non-stationary
chromatic transients in PSR~J1713+0747.  They do not discard the bin.
They re-measure it, and the power there falls without vanishing.  Those
re-measured free-spectral densities have not been released, and the
corrected value is therefore unavailable to us.  We exclude the bin
instead.  This is a response to a datum known to be biased high by an
amount we cannot determine, not an implementation of the reanalysis,
and the bin matters here out of proportion to its number: it is the
only one above the fourth whose $68\%$ interval excludes zero.  In
the published representation it alone sets the shape of the spectrum
above $8$\,nHz.}

\chg{Our fiducial data set is therefore the published
representation~\cite{ng15_kde_zenodo} with one bin removed.  It is
neither the published analysis nor the chromatic one, and two
corrections of opposite sign separate it from the latter.  The
thirteen retained bins still carry the superseded amplitude
normalization, which Ref.~\cite{Agarwal:2026dlv} finds $12\%$ too high;
since $\Ayr \propto 10^{5\dkh/6}$, that residual is $-0.07$\,dex on
$\dkh$, inside the posterior width of Sec.~\ref{sec:res_params}.
Running the other way, discarding the eighth bin removes more than the
reanalysis does, because the corrected bin retains some power, and that
loss pushes $\dkh$ up rather than down.  Neither term can be evaluated
without the unreleased products.  We therefore quote the fiducial
number without claiming that it brackets a full chromatic-noise refit
from either side.}

\chg{Fitting pure power laws through Eq.~\eqref{eq:like} with flat
priors in $(\log_{10}A, \gamma)$ shows both that the implementation is
sound and how far the one-bin edit carries.  Retaining all 14 bins
reproduces the published constraints: with $\gamma = 13/3$ fixed we
obtain $\Ayr = \AyrPLaspub$ (median and $90\%$ interval; published
$\AyrDMX$) and a free power law gives $\gamma = \gammaPLaspub$
(published $\gammaDMX$)~\cite{NANOGrav:2023gor}.  Excluding the eighth bin moves
these to $\Ayr = \AyrPLfid$ and $\gamma = \gammaPLfid$, that is,
\capturedamp\ of the amplitude reduction and \capturedgam\ of the slope
steepening that the full chromatic reanalysis reports
($\AyrCNM$ and $\gamma = \gammaCNM$).
The edit therefore moves the likelihood demonstrably toward the
reanalysis, not somewhere else.}

\chg{Two remarks before the results.  First, what the exclusion costs
and what it leaves untouched is quantified in
Sec.~\ref{sec:systematics}, after the fit itself has been presented.
Second, the critical mass ratio, the stalled fractions, and the
wandering mass budget of Secs.~\ref{sec:stall}
and~\ref{sec:res_stall} are properties of the delay dynamics alone:
they are computed before any likelihood is formed and are unaffected
by this choice, or indeed by the normalization the data prefer.
Independently of all this, the astrophysical reading adopted here is in
better shape than when the signal was announced: $\gamma$ has moved
toward the $13/3$ of circular, gravitational-wave-driven inspiral, and
big-bang nucleosynthesis now excludes the loud end of the cosmological
interpretations~\cite{Elahi:2026ldn}.}

\subsection{Parameters, priors, and sampling}
\label{sec:priors}

\begin{table}
\caption{The five sampled parameters and their priors.  All priors
are flat over the quoted ranges.\label{tab:params}}
\setlength{\tabcolsep}{4pt}
\begin{ruledtabular}
\begin{tabular}{l l l}
parameter & meaning & prior \\
\hline
$\dkh$ & $\Mbh$ normalization offset [dex] & $(-1, 1)$ \\
$\sdex$ & lognormal $\Mbh$ scatter [dex] & $(0.02, 0.8)$ \\
$\thard$ & residual hardening delay [Gyr] & $(0, 6)$ \\
$\et$ & eccentricity at $f_t$ & $(0, 0.9)$ \\
$\log_{10}\rhob$ & nuclear density factor & $(-1.5, 1.5)$ \\
\end{tabular}
\end{ruledtabular}
\end{table}

The model has five parameters, listed with their priors in
Table~\ref{tab:params}.  The first two set the black-hole masses
(Sec.~\ref{sec:painting}).  The delay $\thard$ is the residual
hardening time introduced in Sec.~\ref{sec:thard}: it advances the
coalescence epoch of every pair that survives dynamical friction and
removes those that fail to coalesce by $z=0$, thereby lowering the
amplitude, while $\et$ and $\rhob$ reshape the low-frequency
spectrum through the per-binary turnover
(Secs.~\ref{sec:hardening} and \ref{sec:ecc}).
Everything else (the merger kernel, mass histories, stripping law,
\rev{and the sinking rate}) is fixed by its own calibration, not fit
to the PTA data.  The population weights, base black-hole masses, and
galaxy-merger epochs are independent of $\bm{\theta}$ and are therefore
computed once and cached.  A full likelihood evaluation then costs
$43$\,ms on a single core of an Apple M1 Max, the machine on which all
chains in this paper were run.

We sample with an affine-invariant ensemble~\cite{Goodman:2010,ForemanMackey:2012ig} of 64 walkers for 12000
steps, discarding five autocorrelation times as burn-in.  Acceptance
fractions are $\simeq 0.3$, and effective sample sizes are
$\sim$$2.3\times10^3$ for the prior-edge-truncated $\et$ and
$3.4$--$4.2\times10^3$ for the remaining parameters, sufficient for
the percentile-level statements quoted below.  \chg{The robustness
refits of Sec.~\ref{sec:systematics} use 32 walkers for 6000 steps and
exclude the same frequency bin as the fiducial.}  Identical runs are
performed for the both BK08 and tidal delay models, the latter with the
stall physics active.

\section{Results}
\label{sec:results}

\begin{figure*}
\includegraphics[width=0.9\textwidth]{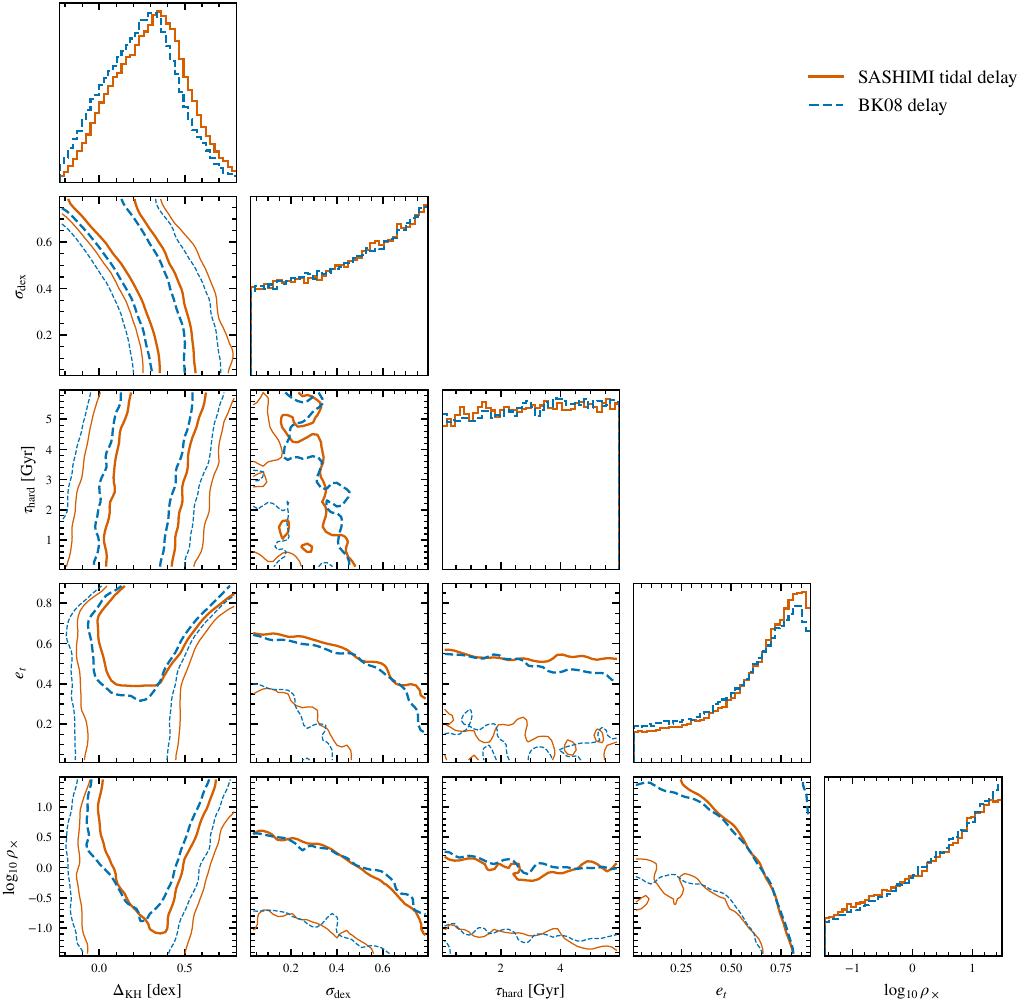}
\caption{Posteriors of the five population parameters from the
NANOGrav 15 yr free spectrum, under \rev{the \sashimi\ tidal delay
(vermilion, solid) and the fixed-mass BK08 delay (blue, dashed)}.
Contours enclose 68\% and 95\%.  The two delay models yield nearly identical constraints on the
sampled parameters (the data cannot distinguish them spectrally) but
imply different underlying populations (Sec.~\ref{sec:res_stall}).
}
\label{fig:corner}
\end{figure*}

\begin{figure}
\includegraphics[width=\columnwidth]{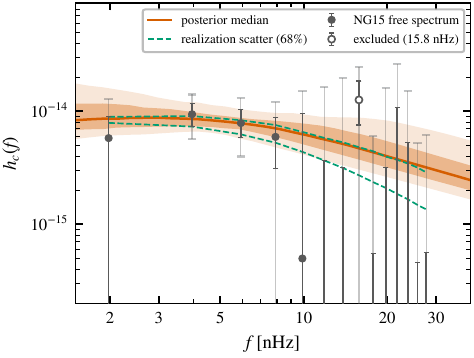}
\caption{Posterior-predictive characteristic strain for the tidal
delay model (median, 68\%, and 95\% bands from 200 posterior draws)
against the NANOGrav 15 yr free-spectrum posteriors.  Dashed lines
mark the 68\% realization (finite-source) scatter of the model,
computed at the posterior median in the circular limit (which
maximizes the single-bin variance) and shown as a fractional
envelope around the median spectrum
(Sec.~\ref{sec:systematics}).  It is distinct from
the parameter-uncertainty bands.  Data points as
in Fig.~\ref{fig:spectrum}.}
\label{fig:ppc}
\end{figure}

\subsection{Population parameters}
\label{sec:res_params}

Figure~\ref{fig:corner} shows the joint posteriors.  The
posterior-predictive spectra in Fig.~\ref{fig:ppc} confirm that the
model family reproduces both the amplitude and the shape of the
measured spectrum.  Marginalized constraints (median and 68\%
interval) are, for the tidal delay model,
\begin{align}
  \dkh &= \dkhT\ {\rm dex}, &
  \sdex &= \sdexT\ {\rm dex}, \nonumber\\
  \thard &= \thardT\ {\rm Gyr}, &
  \et &= \etT, \nonumber\\
  \log_{10}\rhob &= \lrhobT, &
  \label{eq:posteriors}
\end{align}
with the BK08 delay giving statistically indistinguishable values
($\dkh = \dkhB$, $\et = \etB$).  Three features stand out.

First, the black-hole mass normalization is pulled above the
Kormendy--Ho value: $\dkh > 0$ for \pdkhT\ of the posterior mass
(\dkhsigT\ in one-sided Gaussian-equivalent terms).  \chg{This is a
preference, not a detection of a non-zero offset, and it is
deliberately weaker than the published free spectrum would give:
retaining the $15.8$\,nHz bin excluded in Sec.~\ref{sec:likelihood}
raises the median to $\dkh = \dkhaspub$\,dex and the posterior mass
above zero to \pdkhaspub, a shift of $\dkhshiftaspub$\,dex produced by
one frequency bin that the data provider now attributes to pulsar
noise.  We quote the smaller number.}  Recalling
the exact degeneracy of Sec.~\ref{sec:painting}, what is being
measured is $f_{\rm occ}^2\langle\mathcal{M}_c^{5/3}\rangle$.
Since $f_{\rm occ} \le 1$, the data demand \emph{at least} this much
mass excess.  This is the same tension (the nHz sky is loud
relative to local black-hole calibrations) identified in
Refs.~\cite{SatoPolito:2023gym,Simon:2023khd}, anticipated by
merger-driven assembly arguments~\cite{McWilliams:2012an}, and
visible in the
\holodeck\ posteriors~\cite{NANOGrav:2023hfp}.  Its size is under
active revision: recent assessments place it at the $1$--$2.5\sigma$
level or regard it as largely
resolved~\cite{Buttigieg:2026xza,Mingarelli:2026pur}, and an evolving
mass scatter reproduces the data at fixed
normalization~\cite{Matt:2026uyl}, whereas a selection-debiased local
relation, lower by $0.4$--$0.6$\,dex~\cite{Shankar:2016yrf}, would
enlarge the offset instead.  We therefore quote $\dkh$ as a consistency
result rather than a detection.  Here, it emerges from
an independently calibrated merger rate, with delays and stalling
treated physically, which closes the loophole that an underestimated
merger-kernel normalization might mimic the effect (the remaining
rate-level systematics are quantified in
Sec.~\ref{sec:systematics}).  The excess can equally be read
as a preference for larger scatter (the
$\dkh$--$\sdex$ degeneracy in Fig.~\ref{fig:corner}): scatter raises
$\langle\mathcal{M}_c^{5/3}\rangle$ through the heavy tail, as
emphasized by Ref.~\cite{SatoPolito:2023gym}.  

Second, the spectral shape drives a preference for eccentric
binaries and/or dense nuclear environments: the posteriors of $\et$
and $\rhob$ both accumulate toward the upper ends of their prior
ranges and are anticorrelated.  Both parameters bend the
low-frequency spectrum, and \chg{the free fit's slope,
$\gamma = \gammaCNM$ after the chromatic-noise revision, is
accommodated by either of them.  It no longer demands either, because
$13/3$ now lies inside its $95\%$ region.}
Because the $\et$ posterior is truncated by the prior bound
$\et \le 0.9$ (the validity limit we impose on the harmonic
expansion, a numerical rather than physical boundary), the
constraint should be read as a lower bound on the transition
eccentricity rather than a detection of a peak.  \chg{The reading is weaker still for
$\rhob$.  Its $68\%$ interval contains $\log_{10}\rhob = 0$, that is,
no density boost at all, both in the fiducial fit and with the
excluded bin retained; the data are compatible with dense nuclear
environments but at no point require them.  The honest summary of this
feature is that the spectrum does not demand circular,
gravitational-wave-driven binaries, rather than that it demands
eccentricity or dense environments.}
This
echoes the eccentricity preference found by
Ref.~\cite{Raidal:2024odr} with a catalog-based method (their
inferred eccentricity at binary formation is likewise large), and the
same dead end is reached from observed stellar mass
deficits~\cite{Harris:2026mnt}.

Third, the hardening delay $\thard$ is barely constrained: its $68\%$
interval, $\thardT$\,Gyr, spans most of the prior.  It is often assumed
that $\thard$ trades directly against $\dkh$, since longer delays
remove high-redshift mergers and lower the amplitude, which extra
black-hole mass can compensate.  In our posterior, that trade is real
but weak, with a correlation coefficient of only $\rdkhtau$.  The
degeneracy that actually limits $\dkh$ is the one with the mass scatter
$\sdex$ ($\rdkhsig$), followed by $\et$ ($\rdkhet$).  The reason is
that $\thard$ moves only the coalescence epochs, and the pairs whose
epochs it can push past $z=0$ are the high-redshift, subdominant ones,
whereas $\sdex$ rescales $\langle \mathcal{M}_c^{5/3} \rangle$ for
every binary in the integral.  A PTA amplitude measurement alone
therefore constrains the black-hole mass scale up to the mass-scatter
degeneracy, not up to the delay.  The statement is specific to
amplitude-only inference; joint fits that add continuous-wave
non-detections break the same degeneracy from
outside~\cite{Gardiner:2026orz}.

\subsection{The stalled population}
\label{sec:res_stall}

Because the two delay models fit the spectrum equally well, the PTA
data alone do not (yet) distinguish a universe in which small
satellites merge promptly from one in which they linger far longer.
The tidal
and fixed-mass delays differ by \dAdelayabs\ in amplitude, which
$\dkh$ absorbs (their posterior medians differ by $\dkhdiffdel$\,dex).
The spectral fit itself therefore barely tests the delay model. 
The two
models do, however, differ in their electromagnetic predictions.

\begin{table}
\caption{\rev{Wandering-black-hole census under the two delay models.
Every entry is a $z=0$ quantity, and none depends on a sampled parameter.
The upper block counts \emph{satellites}, first summed over all host
masses and then for the single $10^{13}\,\Msun$ host drawn in
Fig.~\ref{fig:stall}; the lower block is a share of the black-hole
\emph{mass} budget.  Every entry carries its own aperture, and entries
from different blocks are not comparable with one another.}}
\label{tab:census}
\begin{ruledtabular}
\begin{tabular}{lcc}
 & tidal & BK08 \\
\colrule
\multicolumn{3}{l}{Accreted satellites not at the nucleus by $z=0$}\\
\multicolumn{3}{l}{\quad\emph{summed over all host masses}}\\
\qquad all & \fstallpure & \fsatbk \\
\qquad stalled (still owing $>t_H$) & \fstallperm & \fstallbk \\
\qquad merely still in transit & \fstalltrans & \fstalltransbk \\
\multicolumn{3}{l}{\quad\emph{in one $10^{13}\,\Msun$ host}
 (Fig.~\ref{fig:stall})}\\
\qquad all & \fstallpureth & \fsatbkth \\
\qquad stalled (still owing $>t_H$) & \fstallpermth & \fstallbkth \\
\qquad merely still in transit & \fstalltransth & \fstalltransbkth \\
\colrule
\multicolumn{3}{l}{Wandering share of the black-hole mass}\\
\quad galaxy scale, $M_{200c}<10^{13}\,\Msun$ & \fwandgal & \fwandgalbk \\
\quad in a $10^{13}\,\Msun$ host & \fwandthirteen & \fwandthirteenbk \\
\quad all hosts & \fwandcos & \fwandcosbk \\
\end{tabular}
\end{ruledtabular}
\end{table}

\rev{Table~\ref{tab:census} collects the census, and it is worth keeping
its two blocks apart.  The upper block counts satellites.  Summed over all
host masses, \fstallpure\ of accreted satellites (with both galaxies
above $10^8\,\Msun$) are not at the nucleus by $z=0$ under the tidal
delay, against \fsatbk\ under the fixed-mass one.  \rev{These are
cosmological averages, dominated by the abundant low-mass hosts.  A single
massive host strands a far larger share: in the $10^{13}\,\Msun$ slice of
Fig.~\ref{fig:stall} the tidal delay leaves \fstallpureth\ of accreted
satellites away from the nucleus, \fstallpermth\ of them stalled
outright, against \fsatbkth\ and \fstallbkth\ for the fixed-mass one.}  Each total divides into
those still owing more than a Hubble time of sinking, the stall proper of
Sec.~\ref{sec:stall}, concentrated below $\xi \simeq \xiperm$, and those
merely still in transit.  The lower block is a share of black-hole
\emph{mass} and varies strongly with host mass, from \fwandgal\ at
galaxy scale to \fwandthirteen\ in a $10^{13}\,\Msun$ host.

One further number, which is not in the table, answers a different
question again, and it is the one the conclusions carry.  Weighting over the
posterior and counting also the nuclear binaries whose galaxies have
merged but whose hardening phase $\thard$ is unfinished, \fnotcoalT\ of
accreted satellites have not produced a coalesced black-hole binary by
$z=0$.  It exceeds the \fstallpure\ of Table~\ref{tab:census} because
it asks whether the black holes have \emph{merged}, not where they
sit.

Hydrodynamic calibrations of the same stage report much shorter
coalescence times~\cite{Francois:2026sca}, and PTA fits prefer short
binary lifetimes~\cite{Gardiner:2026orz}, which appears to contradict
these fractions.  It does not.  Those measurements condition on galaxy
mergers that completed, and simulations that reposition black holes to
galaxy centers at every step~\cite{Weller:2026tbh} exclude the stalled
population from the denominator by construction.  A related constraint
runs the other way, since matching the background amplitude has also
been argued to require central black holes in nearly all galaxies above
$10^9\,\Msun$~\cite{Bonoli:2025qml}; our stalled satellites sit below
that stellar mass, and the two statements are therefore compatible.}
These systems should appear as
long-lived kpc-scale offset nuclei and wandering black holes~\cite{Tremmel:2018wsb,Ricarte:2021qlk,Barausse:2020mdt,IzquierdoVillalba:2020hfk}, with
abundances and host-mass distributions that the framework predicts
deterministically: a parameter-free, falsifiable prediction of the
calibrated delay dynamics, independent of the GWB fit, rather than a
free knob.

\begin{figure}
\includegraphics[width=\columnwidth]{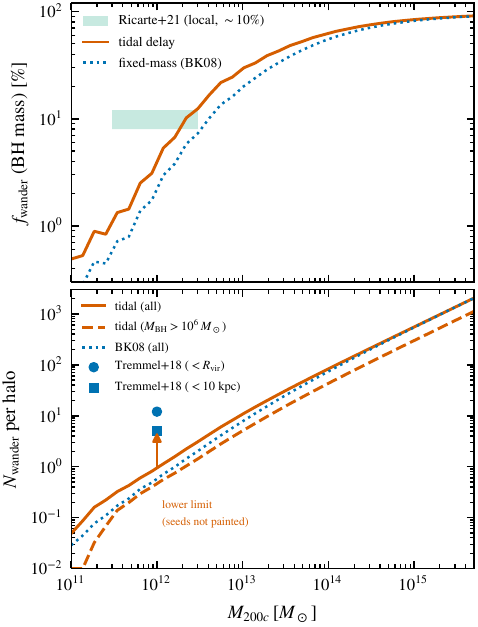}
\caption{Wandering-black-hole census (satellites that do not reach
the nucleus by $z=0$) as a
function of host halo mass, for the tidal (solid) and fixed-mass BK08
(dotted) delays.  Top: the fraction of black-hole mass that is
wandering, against the $\sim$$10\%$ local budget of the Romulus
simulation~\cite{Ricarte:2021qlk}\rev{, whose band is drawn only across
the galaxy-scale decade at which it was measured}.  The two delays diverge most at
galaxy and group scales (where the tidal coupling lifts the
wandering budget by roughly half at group scale and more at
galaxy scale) and converge in clusters, where
both leave the black-hole mass wandering-dominated.  Bottom: the
number of wanderers
per halo (tidal all, and above $10^6\,\Msun$; BK08 all), against the
Milky-Way-mass counts of Ref.~\cite{Tremmel:2018wsb}.  \rev{The upward arrow marks the ``all'' curve, which omits the seed
population below the $M_\star>10^8\,\Msun$ occupation floor and can only
rise.  The count above $10^6\,\Msun$ carries no arrow: the Kormendy--Ho
painting needs $M_\star \simeq 5\times10^{8}\,\Msun$ to reach that mass,
comfortably above the floor, leaving it complete within the model
(Sec.~\ref{sec:res_stall}).  The simulation values are measurements.}
}
\label{fig:wander}
\end{figure}

\rev{Wandering black holes have been tracked before~\cite{Tremmel:2018wsb,Ricarte:2021qlk,IzquierdoVillalba:2020hfk,Untzaga:2024gvk}.}
Our prediction can be confronted with the wandering-black-hole
demographics of cosmological hydrodynamic simulations
(Fig.~\ref{fig:wander}).  Assigning each stalled satellite its
infall black-hole mass and each host its $z=0$ nuclear mass from the
same scaling relation, the framework predicts that wandering black
holes hold a fraction \fwandgal\ of the total black-hole mass in
galaxy-scale halos ($M_{200c} \lesssim 10^{13}\,\Msun$), rising to
\fwandcos\ once cluster hosts are included.  \rev{Romulus finds
$\sim$$10\%$ of the local budget wandering~\cite{Ricarte:2021qlk}, which
our \fwandgal\ matches with no tuning.  That measurement counts
seed-mass black holes, which we do not paint, and our figure is therefore
a lower limit on the same quantity.  Reference~\cite{Ricarte:2021qlk}
further reports that the \emph{number} of wanderers grows roughly
linearly with halo mass, and the like-for-like comparison is our number
slope, $N_{\rm wander} \propto M_{200c}^{\fwandslopeN}$ over
$10^{12}$--$10^{14}\,\Msun$.  The wandering \emph{mass} grows a little
faster, $\propto M_{200c}^{\fwandslope}$, which is why the mass share in
Fig.~\ref{fig:wander} climbs with host mass instead of staying flat.  The
shaded band there marks the local measurement alone, drawn across the
galaxy-scale decade where it was made.}

This agreement does not come from the GWB fit: none of the
five sampled parameters enters the stall determination, and the
wandering mass fraction is exactly independent of $\dkh$ (a uniform
shift of the mass normalization cancels in the ratio).  The comparison,
therefore, tests the framework's delay physics independently of the
inference.  \rev{It also mildly favors the tidal coupling.  Every
\emph{total} in Table~\ref{tab:census} is smaller under the fixed-mass
delay: \fsatbk\
against \fstallpure\ by satellite number, and, in the mass budget,
\fwandgalbk\ against \fwandgal\ at galaxy scale
($M_{200c}<10^{13}\,\Msun$, the aperture of the Romulus comparison
above) and \fwandthirteenbk\ against \fwandthirteen\ in a
$10^{13}\,\Msun$ host.  The fixed-mass delay lengthens the sinking time
less for the low-mass-ratio satellites that dominate the wandering
population.  The separation is modest rather than qualitative, which is
why its discriminating power has to be quantified.}
The
model further predicts that cluster black-hole mass becomes
wandering-dominated (\fwandclus\ at $10^{14}\,\Msun$), the crossover
to a $10\%$ budget occurring near $M_{200c} \simeq \fwandcross\,\Msun$.
The two delays diverge most, in relative terms, at galaxy and group
scales (Fig.~\ref{fig:wander}).  A wandering-black-hole or
offset-AGN census in that regime is therefore the sharpest near-term
discriminant between them.  

Occupation measured in {\sc astrid} now resolves this comparison by
seed model~\cite{Weller:2026occ,Bhowmick:2026brh}, and a volumetric
wandering density anchored to off-nuclear transients is
available~\cite{Guolo:2026oge}.  Two biases act on the mass shares above in
opposite directions, and we do not attempt to net them: not painting
seed-mass black holes makes our share an underestimate, as noted above,
while assigning wanderers the same scaling relation as central black
holes makes it an overestimate, since wanderers descend from stripped
satellites and are undermassive for their
hosts~\cite{Weller:2026tbh}.

Among the massive black
holes ($\Mbh \gtrsim 10^6\,\Msun$) associated with a
$M_{200c} \sim 10^{12}\,\Msun$ host, the tidal coupling leaves a larger
offset (wandering) fraction than the fixed-mass delay.  Classifying
the nuclear-versus-offset status of $\sim$$\Ndisc$ such black holes
separates the two at $3\sigma$ (binomial), or $\sim$$\Ndiscvol$ in a
blind volume-limited sample.  Raising the observability threshold to
$\Mbh \gtrsim 10^7\,\Msun$, safely above the seed floor, gives
$\sim$$\Ndischi$, leaving the forecast insensitive to the seed model.
Individual offset black holes of exactly this kind are now being
localized~\cite{Sfaradi:2026moa,Li:2026wbh}, with complementary
selection from off-nuclear transients~\cite{Guolo:2026oge} and from
spatial offsets at high redshift~\cite{Thakurdesai:2026hwb}.
In clusters, the two delays converge, and a wandering-dominated
black-hole budget (hundreds of wanderers per halo) is a robust,
delay-model-independent prediction of the framework.
The one demographic the framework does \emph{not} reproduce is the
\emph{number} of wanderers in a Milky-Way-mass halo: \Nwandmw\ above
$10^6\,\Msun$, against $\sim$$12$ within the virial radius in
Ref.~\cite{Tremmel:2018wsb}.  \rev{The gap is a seeding difference rather
than a dynamical one.  Both counts apply the same mass cut, but the two
models place $10^6\,\Msun$ black holes in very different galaxies.
Romulus seeds them in objects far below our occupation floor, whereas
painting the Kormendy--Ho relation onto a satellite requires
$M_\star \simeq 5\times10^{8}\,\Msun$ before it carries $10^6\,\Msun$ at
all.  Our census is therefore blind to precisely the seeding physics that
the PTA band cannot constrain either (Sec.~\ref{sec:painting}) and that
the LISA band does.  Above $10^6\,\Msun$ the model count is complete,
while the count of \emph{all} wanderers is the lower limit, marked as
such in Fig.~\ref{fig:wander}.  The mass budget, set by the massive
wanderers the painting does resolve, is the meaningful comparison, and it
is the one that agrees.}

\section{Discussion}
\label{sec:discussion}

\subsection{Systematics budget}
\label{sec:systematics}

Table~\ref{tab:syst} collects the known systematics of the framework,
each quantified by direct numerical experiment
(Appendix~\ref{app:valid}).  The dominant modeling choices (the
merger kernel and the growth of the primary black hole between infall
and coalescence) shift the amplitude at the few-percent to
$\sim$$10\%$ level, comfortably inside the current measurement
uncertainty but relevant for the next data releases.  Our fiducial model
lets the primary black hole co-evolve with its host: because the host
halo keeps accreting and merging after the satellite falls in, we grow
its central black hole along the main branch to the galaxy-merger
epoch, while the satellite (quenched and tidally stripped once it
enters the host) retains its infall mass.  Freezing both masses at
infall instead lowers the amplitude by \dAfrozenabs\ and would pull
$\dkh$ \emph{up} by $\sim$$\dkhfrozen$\,dex.  The excess is, therefore,
robust to this modeling choice: the more conservative frozen
assumption only makes it larger.  As a direct end-to-end
check of the kernel systematic at the inference level, re-running
the entire MCMC with the raw \rev{LC93} kernel shifts the $\dkh$
median by $\dkhshiftLC$\,dex, small compared with its posterior
width.

\chg{The largest single entry in this budget, after the delay model, is
not a modeling choice at all but the data: refitting with the
$15.8$\,nHz bin retained, as in the published free spectrum, moves the
$\dkh$ median by $\dkhshiftaspub$\,dex, to $\dkhaspub$, and the
eccentricity from $\etT$ to $\etaspub$.  The hardening delay does not move at all and the
mass scatter by $0.1\sigma$.  That is the expected
pattern: the excluded bin is the only informative one above $8$\,nHz,
and it therefore constrains the high-frequency \emph{shape} and the
overall normalization and nothing else.  Because the bin lifts both the
amplitude and the required flattening, excluding it weakens two of this
paper's own claims at once, which is the direction that makes the
choice defensible rather than convenient.  How this fiducial relates to a full
chromatic-noise refit is bounded but not signed: the superseded
amplitude normalization carried by the retained bins biases $\dkh$ up
by $\sim$$0.07$\,dex, while discarding the eighth bin outright, rather
than at its corrected value, biases it down
(Sec.~\ref{sec:likelihood}).}

The choice of the local calibration itself is tested the same way:
repeating the entire inference with the McConnell--Ma~\cite{McConnell:2012hz} and Marconi--Hunt~\cite{Marconi:2003hj}
relations in place of Eq.~\eqref{eq:kh13} gives offsets
$\dkh^{\rm MM13} = \dkhMM$ and $\dkh^{\rm MH03} = \dkhMH$.  
The excess is positive under all
three calibrations, and the tension is therefore not an artifact of the
Kormendy--Ho zero point.

A constant offset is also not the only reading of the excess: a
redshift-evolving normalization of the $\Mbh$--$M_{\rm bulge}$
relation has been proposed as an alternative resolution of the same
tension~\rev{\cite{Kozhikkal:2023gkt,NANOGrav:2025mbe}, the latter
finding that a relation growing as $(1+z)^{1.04\pm0.5}$ reproduces the
measured amplitude}.
We, therefore, repeat the fit with
$\dkh(z) = \Delta_{{\rm KH},0} + \Delta_{{\rm KH},1}\, z/(1+z)$
[the functional form of the Moster et al.\ redshift dependence,
Eq.~\eqref{eq:moster}], the host term evaluated at its painting
epoch $z_{\rm gal}$ and the satellite's at infall, with a flat prior
$\Delta_{{\rm KH},1} \in (-1.5, 1.5)$.  The data return
$\Delta_{{\rm KH},1} = \dkhzOne$, consistent with no evolution.

\begin{table}
\caption{Systematics of the framework, as fractional shifts of the
GWB amplitude at fixed parameters.\label{tab:syst}}
\begin{ruledtabular}
\begin{tabular}{l c}
source & $\Delta\Ayr/\Ayr$ \\
\hline
merger kernel (Y11 vs F10 $N$-body) & $\dAkernel$ \\
raw \rev{LC93} kernel (for reference) & $\dAlc$ \\
mass-fn.\ fit (Tinker08~\cite{Tinker:2008ff} vs ST) & $\dAtinkfit$ \\
mass-fn.\ fit (Shirasaki+21~\cite{Shirasaki:2021yof} vs ST) & $\dAsiafit$ \\
mass-fn.\ definition ($180\overline{m}$ vs ST virial) & $\dAstbr$ \\
frozen BH masses (vs fiducial co-evolution) & $\dAfrozen$ \\
tidal vs fixed-mass delay (fiducial) & $\dAdelay$ \\
galaxy-merger threshold $f_{\rm merge} \in (0.02, 0.1)$ & $^{\dAfmhi}_{\dAfmlo}$ \\
orbit-circularity marginalization & $\le \dAorbit$ \\
grid resolution (doubling all axes) & $0.1\%$ \\
stripping integrator vs reference ODE & $0.002\%$ \\
\end{tabular}
\end{ruledtabular}
\end{table}

Beyond these quantified items, five structural assumptions deserve
notice.  (i) We identify bulge mass with total stellar mass.  At the
massive end this is a $\lesssim 0.1$\,dex effect degenerate with
$\dkh$.  A morphology-dependent slope at the high-mass
end~\cite{Graham:2026mds} and the cross-matching of black-hole
catalogs~\cite{Sato-Polito:2025dqw} act on the shape of the calibration
rather than on its normalization and are not captured by the three
swaps below.  Feedback prescriptions move predicted amplitudes by
factors of a few~\cite{Tillman:2026xdi}; fitting $\dkh$ rather than
importing a mass function insulates us from that.  (ii) The mean main-branch history replaces the distribution
of histories.  Assembly variance enters the signal only through the
weak host-mass dependence of the stripping coefficients and the mass
function, and $\pm10\%$ shifts of the assembly history move
population-level rates at the tens-of-percent level only in the
LISA-band extrapolation, not in the PTA band.  (iii) The
Y11 kernel is calibrated on the same simulation suites as
the mass histories.  Kernel and history systematics are, therefore, not
fully independent, and the F10 cross-check bounds their combined
effect.
(iv) Both the sinking march and the nuclear density assignment
idealize their hosts as singular isothermal spheres.  A cored or NFW
host would mainly rescale the sinking rate, \rev{and profile-dependent
residuals in the stall boundary are not quantified here.}  The singular
isothermal sphere is cuspy and has no plateau in its distribution
function, and the core stalling and buoyancy identified for cored
hosts~\cite{Dattathri:2025nac} are therefore absent by construction,
while triaxiality leaves population medians essentially
unchanged~\cite{Chiang:2026txy}.  What we claim is accordingly the
\emph{shape} of the boundary, the tilt $s$, rather than its location.  (v) Sequential mergers are not modeled: a later
satellite can reach the nucleus while an earlier companion is still
stalled, forming a triple system whose Kozai--Lidov cycles and chaotic
exchanges drive a fraction of such pairs to
coalescence~\cite{Bonetti:2017lnj,Li:2026rmc}.  Part of the stalled population of
Sec.~\ref{sec:res_stall} could be rescued through this channel.  \rev{The
quoted stalled fraction is therefore an upper limit on the surviving
wandering population.  Rescued pairs do coalesce, which raises the
background rather than lowering it, and the effect works \emph{against}
the mass excess: a louder model needs less $\dkh$ to match the same data.
The size of that shift is bounded.  Replacing the tidal delay by the
fixed-mass one, which strands fewer satellites, raises $\Ayr$ by
$\dAdelayup$, and additionally setting $\thard = 0$, so that every pair
coalesces as early as the model permits, raises it by \dAearliest.
Since $\Ayr \propto 10^{5\dkh/6}$, even that extreme limit would move
$\dkh$ down by only $\dkhearly$\,dex, under a third of its posterior
width.}

Finally, the likelihood of Sec.~\ref{sec:likelihood} compares the
deterministic ensemble-mean spectrum to the measured per-bin
posteriors, neglecting realization variance: the observed sky is a
single Poisson draw of discrete binaries, and the loudest
contributors are rare~\cite{SatoPolito:2023gym,Sato-Polito:2024lew}.
To quantify this, we decompose the cached population into per-bin
expected source counts and sky-averaged single-source strains (whose
product reproduces the integrand of Eq.~\eqref{eq:phinney} exactly)
and Poisson-sample the realized $\hc^2$ in each of the \nbinfid\ fitted bins at
the posterior-median parameters (Fig.~\ref{fig:ppc}).  The sampling
is performed in the circular limit: an eccentric binary spreads the
same power over many harmonics, which can only decrease the single-bin
variance.  The quoted envelope is therefore conservative.  The 68\%
realization scatter grows from \cvscatlow\ dex in the lowest bin to
\cvscathigh\ dex in the highest, reflecting the decreasing number of
effective sources per bin, but never exceeds \cvratiomax\ of the
corresponding bin's 68\% measurement half-width (the maximum occurs
in bin \cvbinmax).  Added in quadrature, it would inflate the
effective per-bin uncertainty by at most \cvinflmax.  The statement is
about scatter alone: windowing, inter-source interference, and the
offset between the mean and the median spectrum are not modeled, and
the moments require the low-redshift cutoff implicit in our
grid~\cite{Hisamatsu:2026qvd,Zaldarriaga:2026ubp}.  With that caveat,
neglecting realization variance is safe at current
precision~\cite{Raidal:2026ezm,Falxa:2026mhh}, although the full
non-Gaussian treatment will be needed for IPTA-class
data~\cite{Lamb:2025niq,Yu:2026rba,Lin:2026qvc}.

\subsection{Relation to previous work}
\label{sec:previous}

\rev{The framework occupies a deliberate middle ground, and it is worth
separating what is new here from what is inherited.

Three routes to a PTA population model are already in use.  Tree- and
catalog-based pipelines~\cite{Sesana:2008mz,Sesana:2012ak,Kelley:2017lek,
IzquierdoVillalba:2021prf,NANOGrav:2023hfp} carry individual merger
histories and are too slow to sample, reaching the data through a model
bank or a Gaussian-process emulator instead.  A second route parametrizes
the merger rate directly from observed galaxy statistics and bypasses
halos entirely~\cite{Middleton:2015oda,Middleton:2020asl,Chen:2016zyo,Chen:2018znx},
which is fast and has supported emulator-free inference for a decade.  A
third evaluates an analytic EPS halo rate but compresses the whole
halo-merger-to-coalescence stage into one constant
efficiency~\cite{Ellis:2023dgf,Ellis:2023owy,Raidal:2024odr,Ellis:2025dpw,Raidal:2026yef}.
Several groups have since placed a binary population directly inside the
PTA likelihood~\cite{Goncharov:2026cdh,Yu:2026rba,Iemoto:2026fgk,Tomson:2026vwg},
in each case carrying the halo-to-coalescence delay as a fitted
parameter.  What is
new here is not emulator-free sampling, which the second and third routes
already achieve, but that the sampled chain starts at an
$N$-body-calibrated halo kernel and carries an integrated, mass-ratio
dependent delay: against the raw \rev{LC93} kernel the calibration
removes a \dAlcabs\ amplitude bias, and the constant merger probability is
replaced by delay physics that predicts the non-merging fraction rather
than assuming it.

The individual ingredients are largely borrowed, deliberately.  Coupling
tidal stripping to dynamical friction along the orbit is standard in
satellite dynamics~\cite{Jiang:2020rdj,Boldrini:2026thz}, and
simulation-calibrated merger-time prescriptions aimed at pulsar timing
exist~\cite{Francois:2026sca}; what is new is importing that
treatment into a nHz-background calculation, calibrated against the
same simulations that fix \sashimi's subhalo statistics.  That the
background amplitude is far more sensitive to the hardening stage than
to the outer delay is found independently in
Ref.~\cite{Blecha:2026shy}.  The collapse of
the eccentric, environmentally hardened per-binary spectrum onto one
rescaled reference curve is due to Ref.~\cite{Chen:2016zyo}.  Wandering
black holes have been predicted before, both in hydrodynamic
simulations~\cite{Tremmel:2018wsb,Ricarte:2021qlk} and in semi-analytic
models~\cite{IzquierdoVillalba:2020hfk,Untzaga:2024gvk}; what this census
adds is that it is fixed by the same calibrated dynamics that set the
background, with no free normalization.  Among delay treatments the
closest is Ref.~\cite{IzquierdoVillalba:2021prf}, which sets a
stripping-informed starting radius for a subsequent fixed-mass phase.  Our
eccentricity conclusions agree with Ref.~\cite{Raidal:2024odr}.}

\subsection{Outlook: one framework from PTA to LISA}
\label{sec:lisa}

Nothing in Secs.~\ref{sec:framework}--\ref{sec:delays} is specific to
the PTA band.  Extending the mass grid downward and attaching a seed
black-hole occupation function $f_{\rm occ}(M)$ turns the same
integral into a LISA-band merger-rate calculator~\cite{Klein:2015hvg,Barausse:2012fy,Barausse:2020mdt}, with the delay
physics, including stalling (which matters \emph{more} at the low
masses where LISA sources live), carried over unchanged.

Two
consequences are worth stating.  First, the occupation parameters to
which the PTA band is blind (Sec.~\ref{sec:painting}) become the
dominant observable at LISA, where the rate scales as
$f_{\rm occ}^2$: LISA measures precisely the quantity that PTA data
cannot.  Second, the parameters constrained here ($\dkh$, $\sdex$,
$\thard$, and the delay physics) propagate directly into LISA-band
predictions.  Because $\dkh$ and $\thard$ enter both bands, and because
a cached evaluation costs milliseconds in either, the two data sets can
be sampled \emph{together}, over shared parameters, rather than one
being conditioned on the other.
Joint PTA--LISA population
inference in this framework, together with individual-source and
anisotropy statistics, will be presented elsewhere.

\section{Conclusions}
\label{sec:conclusions}

We have built and tested a population framework for the SMBH-binary
GWB combining three properties that are not found together elsewhere,
the second of which is the one that distinguishes it: it is
merger-tree-free (a deterministic triple
integral over an $N$-body-calibrated EPS accretion kernel,
resolution-free by construction), its merger delays are regulated by
subhalo tidal evolution coupled continuously to dynamical friction
\rev{(predicting a stalled, wandering-black-hole population
below a sharp critical mass ratio $\xi \simeq \xiperm$)}, and it is
fast enough (milliseconds per likelihood) to be sampled directly against
PTA data without emulation.  Direct sampling has since been achieved by
several groups~\cite{Goncharov:2026cdh,Yu:2026rba,Iemoto:2026fgk,Tomson:2026vwg};
what remains particular to this framework is that the delay is
integrated from the stripping and sinking equations rather than fitted.  Applied to the NANOGrav 15 yr
Hellings--Downs free spectrum, \chg{excluding the one bin whose excess power the
data provider has since attributed to pulsar noise,} the fit yields a
black-hole mass normalization $\dkh = \dkhT$\,dex above the
Kormendy--Ho relation~\cite{Kormendy:2013dxa} (degenerate with
occupation and scatter, and robust to the choice of delay model,
merger kernel, local mass calibration, and its redshift evolution)
\chg{and a spectrum compatible with, but not requiring, eccentric
binaries or dense stellar environments.  Retaining the excluded bin
would raise the normalization to $\dkhaspub$\,dex; we prefer the more
conservative reading.}  Independently of the fit, the calibrated
delay dynamics predict a parameter-free stalled fraction
of \fstallpure\ of accreted satellites unmerged by $z=0$ (the
posterior-weighted non-coalesced fraction, adding still-hardening
nuclear binaries, is \fnotcoalT),
whose wandering-black-hole mass
budget matches cosmological simulations without tuning and exceeds
that of a fixed-mass delay.  This is an independent,
electromagnetically testable prediction of the same framework.  The
identical machinery is designed to extend to the
LISA band, where the occupation physics invisible to PTAs becomes
the leading observable.

The code implementing the framework, including the cached-likelihood
evaluator and the analysis chains of this paper, will be made
available with \sashimi.

\begin{acknowledgments}
This work was supported by JSPS KAKENHI Grant Number 24K07039.
This work made use of the NANOGrav 15 yr free-spectrum KDE data products
(Zenodo, doi:10.5281/zenodo.10344086).
\end{acknowledgments}

\appendix
\section{The reduced \rev{LC93} kernel}
\label{app:lc}

For the kernel comparison of Sec.~\ref{sec:kernels}, we need the raw
EPS merger rate in the same per-host convention as
Eq.~\eqref{eq:Na}.  \rev{LC93}~\cite{Lacey:1993iv} gives,
in its Eq.~(2.18), the rate for a halo of mass $M_1$ at cosmic time
$t$ to merge with a companion $\Delta M$, forming
$M_2 = M_1+\Delta M$:
\begin{align}
  \frac{d^2 p}{d\ln \Delta M\, dt}
  ={}& \sqrt{\frac{2}{\pi}}\,
     \frac{1}{t}\left|\frac{d\ln \delta_c}{d\ln t}\right|
     \frac{\Delta M}{M_2}
     \left|\frac{d\ln \sigma_2}{d\ln M_2}\right|
     \nonumber\\
   & \times
     \frac{\delta_c(t)}{\sigma_2}
     \left(1-\frac{\sigma_2^2}{\sigma_1^2}\right)^{-3/2}
     \nonumber\\
   & \times
     \exp\!\left[-\frac{\delta_c^2(t)}{2}
     \left(\sigma_2^{-2}-\sigma_1^{-2}\right)\right],
  \label{eq:lc218}
\end{align}
with $\sigma_i = \sigma(M_i)$ the $z=0$ rms fluctuation and
$\delta_c(t) = 1.686/D(t)$.  Using
$({1}/{t})|d\ln\delta_c/d\ln t| =
|d\delta_c/dt|/\delta_c$ and
$|d\ln\sigma_2/d\ln M_2|/\sigma_2 =
M_2 |dS_2/dM_2|/(2 S_2^{3/2})$ with $S = \sigma^2$, the
$M_2$ and $\delta_c$ factors cancel and Eq.~\eqref{eq:lc218} reduces 
to
\begin{align}
  \frac{d^2 p}{d\ln \Delta M\, dt}
  ={}& \frac{1}{\sqrt{2\pi}}
  \left|\frac{d\delta_c}{dt}\right| \Delta M
  \left|\frac{dS_2}{dM_2}\right| S_2^{-3/2}
  \nonumber\\
  & \times
  \left(1-\frac{S_2}{S_1}\right)^{-3/2}
  \exp\!\left[-\frac{\delta_c^2}{2}
  \left(\frac{1}{S_2}-\frac{1}{S_1}\right)\right].
  \label{eq:lcreduced}
\end{align}
Dividing by $H(z)$ converts the rate per cosmic time to the
per-$\ln(1+z)$ convention of Eq.~\eqref{eq:Na}.  The identification
$M_1 = M(z_a)$, $\Delta M = m_a$, truncated at $m_a \le M_1$, then
makes Eq.~\eqref{eq:lcreduced} a drop-in replacement for the Y11
kernel.  Equation~\eqref{eq:lcreduced} exhibits the well-known
$\Delta M^{-1/2}$ divergence for $\Delta M \ll M_1$ and the EPS
progenitor-exchange asymmetry, both inherited by any calculation
built on it.

\section{Validation and convergence}
\label{app:valid}

We list the numerical validations behind the quoted systematics
(Table~\ref{tab:syst}).  All are reproducible from the public
analysis scripts.

\paragraph{Grid convergence.}  Doubling all three grid dimensions
($36\times48\times64 \to 72\times96\times128$) changes $\Ayr$ by
$0.14\%$.

\paragraph{Stripping integrator.}  In the variable $u = r^2$ the
right-hand side of Eq.~\eqref{eq:sink} is independent of $r$.
Equations~\eqref{eq:strip} and \eqref{eq:sink} therefore integrate as a
single vectorized march in redshift over the whole grid, with no
per-orbit ODE solves.  The march agrees with a reference stiff ODE
integration of Eq.~\eqref{eq:strip} to $0.002\%$ in the final bound
mass, with second-order convergence in step number ($0.07\%$ amplitude
shift from $n_{\rm step} = 250$ to $1000$ at the fiducial parameters,
and we use 500).

\paragraph{Eccentric spectra.}  At $\et = 0$ the harmonic sum
Eq.~\eqref{eq:phi} reproduces the circular suppression factor to
$0.01\%$.  The identity $\sum_n g(n,e) = F(e)$ is satisfied to
$10^{-3}$ over the tabulated range, and the adaptive harmonic cut
captures $99.9\%$ of $F(\et)$ everywhere inside the prior.

\paragraph{Fast evaluator.}  The cached likelihood model
(Sec.~\ref{sec:priors}) matches the direct per-cell harmonic
calculation to $0.07\%$ at all tested eccentricities.

\paragraph{Merger threshold.}  Recomputing the population for
$f_{\rm merge} \in \{0.02, 0.035, 0.05, 0.07, 0.10\}$ gives
$\Ayr$ shifts of $\{\dAfmlist\}$ and \rev{non-coalesced
fractions (at $\thard = 0.5$\,Gyr, and therefore larger than the
galaxy-merger stall of Table~\ref{tab:census}) of $\{\fnotfmlist\}$}: the calibration
absorbs the threshold arbitrariness because the $r$-independent
sinking rate in $u = r^2$ makes the final approach to any threshold
fast.

\paragraph{Orbit circularity.}  Scaling both the sinking clock
[$\propto e^{-1.9(\eta-0.5)}$, following BK08] and the stripping
clock [$\propto e^{-\kappa(\eta-0.5)}$ with $\kappa \simeq 2$--$4$
implied by the orbit dependence of bound-mass evolution in
Refs.~\cite{Ogiya:2019del,Penarrubia:2010jk}] and marginalizing over
the infall circularity distribution
$P(\eta) \propto \eta^{1.22}(1-\eta)^{1.22}$~\cite{Zentner:2004dq}
changes $\Ayr$ by at most \dAorbit\ across
$\kappa \in \{0, 1.9, 3.0, 3.8\}$: faster sinking and faster
stripping on radial orbits nearly cancel.

\paragraph{Mass-function conversion.}  The alternative mass functions
(Sec.~\ref{sec:master}) are calibrated on different halo mass
definitions.  Before evaluating each, we convert the pipeline's
$M_{200c}$ to the fit's native spherical overdensity using the
concentration of Ref.~\cite{Correa:2015cma} and an exact
Navarro--Frenk--White profile: for a target overdensity $\Delta_c$
relative to the critical density we solve $f(y)/y^3 = (\Delta_c/200)\,
f(c)/c^3$ with $f(x) = \ln(1+x)-x/(1+x)$ and $c = c_{200c}(M,z)$, giving
$M_\Delta/M_{200c} = f(y)/f(c)$.  The abundance carries the Jacobian
$d\ln M_\Delta/d\ln M_{200c}$.  Against \sashimi's own
$M_{200c}\!\to\!M_{\rm vir}$ fit this exact solve agrees to $0.4\%$.
The Tinker et al.\ overdensity, defined relative to the mean density,
maps to $M_{200c}$ exactly at $\Delta = 200/\Omega_{\rm m}(z)$, and no
profile conversion is needed there.

\paragraph{Power-law recovery.}  Fitting pure power laws through the
free-spectrum likelihood recovers the published NANOGrav 15 yr
constraints (Sec.~\ref{sec:likelihood}).

\paragraph{Sampler.}  The affine-invariant implementation recovers
the mean of a correlated five-dimensional Gaussian to $0.04\sigma$
and its covariance to $3\%$ in the whitened norm.  On the physical
posteriors, acceptance is $\simeq 0.3$; the two production chains run
$\simeq 40$ autocorrelation times and the shorter robustness refits
$16$--$22$.

\bibliography{refs}

\end{document}